\documentclass[11pt,journal,draftcls,onecolumn]{IEEEtran}
\IEEEoverridecommandlockouts
\usepackage{amsmath, amsthm, amssymb}
\usepackage{verbatim}
\usepackage{graphicx}
\usepackage{cite}
\usepackage{subfig}
\usepackage[subnum]{cases}

\usepackage[usenames]{color}
\definecolor{plum}  {rgb}{.4,0,.4}
\definecolor{BrickRed} {rgb}{0.6,0,0}

\usepackage[plainpages=false,pdfpagelabels,colorlinks=true,linkcolor=BrickRed,citecolor=plum]{hyperref}

\newtheorem{theorem}{Theorem}
\newtheorem{lemma}{Lemma}

\newtheorem{corollary}{Corollary}
\newtheorem{remark}{Remark}
\newtheorem{example}{Example}
\newcommand{\R}{\mathbb{R}}

\newcommand{\E}{\mathbb{E}}

\newcommand{\Var}{{\rm Var}}
\newcommand{\I}{\mathbf{1}}

\def\cL{{\mathcal L}}

\def\cS{{\mathcal S}}

\def\sU{{\mathsf U}}

\def\sW{{\mathsf W}}
\def\sX{{\mathsf X}}
\def\sY{{\mathsf Y}}
\def\sZ{{\mathsf Z}}

\def\PP{{\mathbb P}}
\def\QQ{{\mathbb Q}}

\def\deq{\triangleq}
\def\wh#1{{\widehat{#1}}}

\def\eps{\varepsilon}
\begin{document}
\sloppy

\title{Information-Theoretic 
Lower Bounds on 
Bayes Risk in
Decentralized Estimation
}

\author{Aolin Xu and Maxim Raginsky~\IEEEmembership{Senior Member,~IEEE}
\thanks{This work was supported by the NSF under grant CCF-1017564, CAREER award CCF-1254041, by the Center for Science of
Information (CSoI), an NSF Science and Technology Center, under grant
agreement CCF-0939370, and by ONR under grant N00014-12-1-0998. The material in this paper was presented in part at the IEEE International Symposium on Information Theory (ISIT), Hong Kong,  June 2015.}
\thanks{
The authors are with the Department of Electrical and Computer Engineering and the Coordinated Science Laboratory, University of Illinois, Urbana, IL 61801, USA. E-mail: \{aolinxu2,maxim\}@illinois.edu. }
}
\maketitle


\begin{abstract}
We derive lower bounds on the Bayes risk in decentralized estimation, where the estimator does not have direct access to the random samples generated conditionally on the random parameter of interest, but only to the data received from local processors that observe the samples.
The received data are subject to communication constraints, due to quantization and the noise in the communication channels from the processors to the estimator.
We first derive general lower bounds on the Bayes risk using information-theoretic quantities, such as mutual information, information density, small ball probability, and differential entropy. We then apply these lower bounds to the decentralized case, using strong data processing inequalities to quantify the contraction of information due to communication constraints. We treat the cases of a single processor and of  multiple processors, where the samples observed by different processors may be conditionally dependent given the parameter, for noninteractive and interactive communication protocols.
Our results  recover and improve recent lower bounds on the Bayes risk and  the minimax risk  for certain decentralized estimation problems, where previously only conditionally independent sample sets and noiseless channels have been considered.
Moreover, our results provide a general way to quantify the degradation of estimation performance caused by distributing resources to multiple processors, which is only discussed for specific examples in existing works.

\end{abstract}

\section{Introduction}
\subsection{Decentralized estimation}
In decentralized estimation, the estimator does not have direct access to the samples generated according to the parameter of interest, but only to the data received from local processors that observe the samples. 
In this paper, we consider a general model of decentralized estimation, where each local processor observes a set of samples generated according to a common random parameter $W$, quantizes the samples to a fixed-length binary message, then encodes and sends the message to the estimator over an independent and possibly noisy communication channel. 
When the communication channels are noiseless and feedback from the estimator to the local processors is available, the processors can operate in an interactive protocol by taking turns to send messages, where the message sent by each processor can depend on the previous messages sent by the other processors. An estimate $\wh W$ is then computed based on the messages received from the local processors.
The estimation performance is measured by the expected distortion between $W$ and $\wh W$, with respect to some distortion function.
The minimum possible expected distortion is defined as the Bayes risk.
We derive lower bounds on the Bayes risk for this estimation problem, and gain insight into the fundamental limits of decentralized estimation.

There are three types of constraints inherent in decentralized estimation.
The first, and the most fundamental one, is the \textit{statistical constraint}, determined by the joint distribution of the parameter and the samples.
The statistical constraint exists even in the centralized estimation, where the estimator can directly observe the samples.
To study how the estimation performance is limited by the statistical constraint, we start with deriving lower bounds on the Bayes risk for centralized estimation in Section~\ref{sec:RB_lb_central}.
The results obtained in Section~\ref{sec:RB_lb_central} apply to the decentralized estimation as well, but, more importantly, they also serve as the basis for the refined lower bounds for the decentralized estimation in Section~\ref{sec:single} and Section~\ref{sec:multi}.

The second is the \textit{communication constraint}, due to the separation between the local processors and the estimator. 
The communication constraint arises even when there is only one local processor.
It can be caused by the finite precision of analog-to-digital conversion, limitations on the storage of intermediate results, limited transmission blocklength, channel noise, etc. 
In Section~\ref{sec:single}, we present a detailed study of decentralized estimation with a single processor and reveal the influence of the communication constraint on the estimation performance. Section~\ref{sec:SDPI_mi} contains background information on \textit{strong data processing inequalities}, the major tool used in our analysis of the communication constraint.

The third constraint appears when there are more than one local processors.
It is the \textit{penalty of decentralization}, caused by distributing the samples and communication resources to multiple processors.
We study decentralized estimation with multiple processors in Section~\ref{sec:multi}, where we show that, regardless of whether or not the sample sets seen by different local processors are conditionally independent given the parameter, the degradation of estimation performance becomes more pronounced when the resources are distributed to more processors.
We also provide lower bounds on the Bayes risk for interactive protocols, where the processors take turns to send their messages, and each processor sends one message based on its sample set and the previous messages sent by other processors.

\subsection{Method of analysis}

Our method of analysis is information-theoretic in nature. The major quantity we examine is the conditional mutual information $I(W;\wh W|U)$ with a judiciously chosen auxiliary random variable $U$.

We first lower-bound this quantity according to the estimation performance, such as the  probability of excess distortion or the expected distortion.
The lower bounds will also depend on the \textit{a priori} uncertainty about $W$, measured either by its small ball probability or by its differential entropy.
Any such lower bound can be viewed as a generalization of Fano's inequality, which indicates the least amount of information about $W$ that must be contained in $\wh W$ in order to achieve a certain estimation performance. We also analyze the  probability of excess distortion and the expected distortion via the distribution of the conditional information density $i(W;\wh W|U)$.

On the other hand, various constraints inherent in decentralized estimation impose upper bounds on $I(W;\wh W|U)$.
According to the statistical constraint, $I(W;\wh W|U)$ is upper-bounded by the conditional mutual information between $W$ and the samples.
The communication constraint further implies that the amount of information about $W$ contained in the estimator's indirect observation of the samples will be a contraction of the amount contained in the samples. 
We use strong data processing inequalities to quantify this contraction of information and to couple the communication constraint and the statistical constraint together in the upper bounds on $I(W;\wh W|U)$.
When there are multiple processors, strong data processing inequalities also give an upper bound that decreases as the samples and communication resources are distributed to more processors, which reflects the penalty of decentralization.
In addition, we rely on a cutset analysis that chooses the conditioning random variable $U$ to consist of all the samples seen by only a subset of the processors; this choice is useful for analyzing the situation where the processors observe sample sets that are dependent conditional on $W$.

Finally, by combining the upper and lower bounds on $I(W;\wh W|U)$, we obtain lower bounds on the Bayes risk.

\subsection{Related works}
The early works on the fundamental limits of decentralized estimation mainly focused on the asymptotic setting, e.g., determining the error exponent in multiterminal hypothesis testing with fixed quantization rates. Those works are surveyed by Han and Amari \cite{Han_inf98}.
In recent years, the focus has shifted towards determining explicit dependence of the estimation performance on the communication constraint (see, e.g., \cite{Zhang_dist13,Duchi_dist,GMN14,BGMNW15,Shamir_dist14} and references therein). 
For instance, Zhang et al.~\cite{Zhang_dist13} and Duchi et al.~\cite{Duchi_dist} derived lower bounds on the minimax risk of several decentralized estimation problems with noiseless communication channels. 
Their results also provide lower bounds on the number of bits needed in quantization to achieve the same minimax rate as in the centralized estimation.
Garg et al.~\cite{GMN14} extended the lower bound for interactive protocols in \cite{Zhang_dist13}, which centered on the one-dimensional Gaussian location model, to the setting of high-dimensional Gaussian location models.
Braverman et al.~\cite{BGMNW15} presented lower bounds for decentralized estimation of a sparse multivariate Gaussian mean. Their derivation is based on a ``distributed data processing inequality," which quantifies the information loss in decentralized binary hypothesis testing under the Gaussian location model.
Shamir \cite{Shamir_dist14} showed that the analysis of several decentralized estimation and online learning problems can be reduced to a certain meta-problem involving discrete parameter estimation with interactive protocols, and derived minimax lower bounds for this meta-problem.

The main idea underlying all of the above works is that one has to quantify the contraction of information due to the communication constraint; however, this is often done in a case-by-case manner for each particular problem, and the resulting contraction coefficients are generally not sharp. Additionally, these works only consider the situation where the sample sets are conditionally independent given the parameter and where the communication channels connecting the processors to the estimator are noiseless.

By contrast, we derive general lower bounds on the Bayes risk, which automatically serve as lower bounds on the minimax risk. We use strong data processing inequalities as a unifiying general method for quantifying the contraction of mutual information in decentralized estimation. Our results apply to general priors, sample generating models, and distortion functions.  When particularized to the examples in the existing works, our results can lead to sharper lower bounds on both the Bayes and the minimax risk. For example, we improve the lower bound for the mean estimation on the unit cube studied in \cite{Zhang_dist13}, as well as the lower bound for the meta-problem of Shamir \cite{Shamir_dist14}.
Moreover, we consider the situations where the sample sets are conditionally dependent and where the communication channels are noisy. 
We also provide a general way to quantify the degradation of estimation performance caused by distributing resources to multiple processors, which is only discussed for specific examples in existing works.

\subsection{Notation}

In this paper, all logarithms are binary, unless stated otherwise. 
A vector like $(X_1,\ldots,X_n)$ may be abbreviated as $X^n$.
For $r,s\in\R$, $r \wedge s \deq \min\{r,s\}$. For an integer $m$, $[m] \deq \{1,\ldots,m\}$.
For functions $f$ and $g$, $f(x) \sim g(x)$ means that $\lim_{x\rightarrow\infty}{f(x)}/{g(x)} = 1$, while $f(x) \gtrsim g(x)$ means that $\liminf_{x\rightarrow\infty}{f(x)}/{g(x)} \ge 1$.
We use $h_2(p) \deq -p \log p - (1-p)\log (1-p)$ and $d_2(p \| q) \deq p \log \frac{p}{q} + (1-p)\log \frac{1-p}{1-q}$ to denote the binary entropy and the binary relative entropy functions.

\section{Bayes risk lower bounds for centralized estimation}\label{sec:RB_lb_central}
In the standard Bayesian estimation framework, $\mathcal P = \{P_{X|W=w}: w\in\sW\}$ is a family of distributions on an observation space $\sX$, where the parameter space $\sW$ is endowed with a prior distribution $P_W$.
Given $W=w$, a sample $X$ is generated from $P_{X|W=w}$.
In centralized estimation, the unknown random parameter $W \sim P_W$ is estimated from $X$ as $\wh W = \psi(X)$, via an estimator $\psi:\sX \rightarrow \sW$.
Given a non-negative distortion function $\ell:\sW\times\sW \rightarrow \R^+$, define the Bayes risk for estimating $W$ from $X$ with respect to $\ell$ as
\begin{align}
R_{\rm B} = \inf_{\psi} \E[\ell (W,\psi(X))] .
\end{align}
In this section, we derive lower bounds on the Bayes risk in the context of centralized estimation. These bounds serve as lower bounds for the decentralized setting as well, but they can also be used to derive refined lower bounds for decentralized estimation, as shown in Sections~\ref{sec:single} and \ref{sec:multi}. We first present lower bounds on the Bayes risk based on small ball probability, mutual information, and information density in Sections~\ref{sec:RB_lb_cent_mi} and \ref{sec:RB_lb_cent_id}.
These lower bounds apply to estimation problems with an arbitrary joint distribution $P_{W,X}$ and an arbitrary distortion function $\ell$, and also provide generalizations of Fano's inequality, as discussed in Section~\ref{sec:RB_lb_cent_Fano}.
Next, in Section~\ref{sec:RB_lb_cent_diff}, we present a lower bound based on mutual information and differential entropy, which applies to parameter estimation problems in $\R^d$, with distortion functions of the form $\ell(w,\wh w) = \|w-\wh w\|^r$ for some norm $\|\cdot\|$ and some $r\ge 1$.

\subsection{Lower bounds based on mutual information and small ball probability}\label{sec:RB_lb_cent_mi}

The \textit{small ball probability} of $W$ with respect to distortion function $\ell$ is defined as
\begin{align}
\cL_{W} (\rho) = \sup_{w\in \sW} \PP[\ell (W,w) < \rho] .
\end{align}
Given another random variable $U$ jointly distributed with $W$, the conditional small ball probability of $W$ given $U=u$ is defined as
\begin{align}
\cL_{W|U} (u,\rho) = \sup_{w\in \sW} \PP[\ell (W,w) < \rho | U=u] .
\end{align}
These two quantities measure the spread of $P_W$ or $P_{W|U=u}$, respectively. The smaller the small ball probability, the more spread the corresponding distribution is w.r.t.\ the distortion function $\ell$. We give a lower bound on the probability of excess distortion in terms of conditional mutual information and conditional small ball probability:
\begin{lemma}\label{lm:excess_lb_csbp}
For any estimate $\wh W$ of $W$, any $\rho> 0$, and any auxiliary random variable $U$,
\begin{align}\label{eq:1-pt_lb}
\PP[\ell(W,\wh W) \ge \rho] \ge 1 - \frac{I(W;\wh W | U) + 1}{\log \big(1/\E[ \cL_{W|U}(U,\rho)] \big)} .
\end{align}
\end{lemma}
\begin{IEEEproof}
The inequality \eqref{eq:1-pt_lb} is a direct consequence of the following lower bound on the conditional mutual information obtained in \cite{ISIT14_dist_comp}: whenever $\PP[\ell(W,\wh W) \ge \rho]\le\delta$,
\begin{align*}
I(W;\wh W|U) \ge (1-\delta)\log\frac{1}{\E[ \cL_{W|U}(U,\rho)]} - h_2(\delta) .
\end{align*}
In Appendix~\ref{appd:RB_lb_NP}, we present an alternative unified proof of Lemmas~\ref{lm:excess_lb_csbp} and \ref{lm:excess_lb_iden} using properties of the Neyman--Pearson function.
\end{IEEEproof}
Our first lower bound on the Bayes risk for centralized estimation is an immediate consequence of Lemma~\ref{lm:excess_lb_csbp}:
\begin{theorem}\label{th:RB_lb_mi}
The Bayes risk for estimating the parameter $W$ based on the sample $X$ with respect to the distortion function $\ell$ satisfies
\begin{align}\label{eq:RBlb_condMI}
R_{\rm B} &\ge \sup_{P_{U|W,X}} \sup_{\rho>0} \rho \left(1 - \frac{I(W;X|U) + 1}{\log (1/\E[\cL_{W|U}(U,\rho)])} \right) .
\end{align}
In particular,
\begin{align}\label{eq:RBlb_mi}
R_{\rm B} &\ge \sup_{\rho>0} \rho \left(1 - \frac{I(W;X) + 1}{\log (1/\cL_{W}(\rho))} \right) .
\end{align}
\end{theorem}
\begin{IEEEproof}
For an arbitrary estimator $\psi:\sX\rightarrow\sW$, 
\begin{align}
I(W;\wh W|U) \le I(W;X|U)
\end{align}
by the data processing inequality.
It follows from Lemma~\ref{lm:excess_lb_csbp} that
\begin{align}
\PP[\ell(W,\wh W) \ge \rho] \ge 1 - \frac{I(W;X| U) + 1}{\log \big(1/\E[ \cL_{W|U}(U,\rho)] \big)}, \qquad \rho>0 . 
\end{align}
Theorem~\ref{th:RB_lb_mi} follows from  Markov's inequality
$\E[\ell(W,\wh W)] \ge \rho\, \PP[\ell(W,\wh W) \ge \rho]$ and from the arbitrariness of $\psi$, $P_{U|W,X}$, and $\rho>0$.
\end{IEEEproof}

\begin{remark}\label{rmk:th:RB_lb_mi}
{\em Precise evaluation of the expected conditional small ball probability $\E[\cL_{W|U}(U,\rho)]$ in Theorem~\ref{th:RB_lb_mi} can be difficult. The following technique may sometimes be useful: Suppose we can upper-bound $\E[\cL_{W|U}(U,\rho)]$ by some increasing function $g(\rho)$, which has an inverse function $g^{-1}(p) = \sup\{\rho > 0: g(\rho) \le p\}$. Given some $s \in (0,1)$, choosing a suitable $\rho>0$ such that 
\begin{align}
g(\rho) \le 2^{-(I(W;X | U) + 1)/(1-s)} 
\end{align}
guarantees
\begin{align}
1 - \frac{I(W;X | U) + 1}{\log \big(1/\E[\cL_{W|U}(U,\rho)] \big)} \ge s .
\end{align}
It then follows from Theorem~\ref{th:RB_lb_mi} that
\begin{align}\label{eq:RB_sup_s}
R_{\rm B} &\ge \sup_{P_{U|W,X}}\sup_{0<s<1} s g^{-1}\left( 2^{-({I(W;X | U) + 1})/(1-s)} \right) .
\end{align}}
\end{remark}
A similar methodology for deriving lower bounds on the Bayes risk has been recently proposed by Chen et al.~\cite{Chen_Bayes}, who obtained unconditional lower bounds similar to \eqref{eq:RBlb_mi} in terms of general $f$-informativities \cite{Csiszar72} and a quantity essentially the same as the small ball probability.
However, as will be shown later, the conditional lower bound \eqref{eq:RBlb_condMI} can lead to tighter results compared to the unconditional version \eqref{eq:RBlb_mi}, and is also useful in the context of decentralized estimation problems.

For the problem of estimating $W$ based on $n$ samples $X_1,\ldots,X_n$ conditionally i.i.d.\ given $W$, we can choose the conditioning random variable $U$ in \eqref{eq:RBlb_condMI} to be an independent copy of $X^n = (X_1,\ldots,X_n)$ conditional on $W$, denoted as $X'^n$ --- that is, $P_{W,X^n,X'^n} = P_W \otimes P_{X^n|W} \otimes P_{X'^n|W}$ and $P_{X'^n|W}=P_{X^n|W}$. This choice leads to
\begin{align}\label{eq:RB_lb_X'n}
R_{\rm B} &\ge \sup_{\rho > 0} \rho \left(1 - \frac{I(W ; X^n|X'^n) + 1}{\log\big(1/\E[\cL_{W|X^n}(X^n,\rho)]\big)}\right) .
\end{align}
We then need to evaluate or upper-bound $I(W ; X^n|X'^n)$ and $\E[\cL_{W|X^n}(X^n,\rho)]$. For example, in  the smooth parametric case when $\mathcal P$ is a subset of a finite-dimensional exponential family and $W$ has a density supported on a compact subset of $\R^d$, it was shown by Clarke and Barron \cite{Cla_Bar94,Clarke_Barron} that
\begin{align}\label{eq:I_WXn_ClaBar}
I(W ; X^n) = \frac{d}{2}\log\frac{n}{2\pi e} + h(W) + \frac{1}{2}\E[\log \det J_{X|W}(W)] + o(1)  \qquad \text{as $n \rightarrow \infty$}
\end{align}
where $h(W)$ is the differential entropy of $W$, and $J_{X|W}(w)$ is the Fisher information matrix about $w$ contained in $X$. When  \eqref{eq:I_WXn_ClaBar} holds, we have
\begin{align}
I(W ; X^n|X'^n) &= I(W ; X^n, X'^n) - I(W ; X'^n) \\
&\rightarrow \frac{d}{2}  \qquad \text{as $n \rightarrow \infty$}  \label{eq:IWX|X'_asym}
\end{align}
meaning that $I(W ; X^n|X'^n)$ in \eqref{eq:RB_lb_X'n} is asymptotically independent of $n$.
Upper-bounding $\E[\cL_{W|X^n}(X^n,\rho)]$ is more problem-specific.
We give two examples below, in both of which we consider the absolute distortion $\ell(w,\wh w) = |w-\wh w|$, such that the Bayes risk gives the Minimum Mean Absolute Error (MMAE).
A benefit of lower-bounding MMAE is that the square of the resulting lower bound also serves as a lower bound for the Minimum Mean Squared Error (MMSE).

\begin{example}[Estimating Gaussian mean with Gaussian prior]\label{ex:GaussMean_GaussPrior}
Consider the case where the parameter $W \sim N(0,\sigma_W^2)$, the samples are $X_i = W + Z_i$ with $Z_i \sim N(0,\sigma^2)$ independent of $W$ for $i=1,\ldots,n$, and $\ell(w,\wh w) = |w-\wh w|$.
\end{example}
\noindent From the conditional lower bound \eqref{eq:RB_lb_X'n}, we get the following lower bound for Example~\ref{ex:GaussMean_GaussPrior}:
\begin{corollary}\label{co:GaussMean_GaussPrior}
In Example~\ref{ex:GaussMean_GaussPrior}, the Bayes risk is lower bounded by
\begin{align}
R_{\rm B}&\ge \frac{1}{16} \sqrt{\frac{\pi\sigma_W^2}{2(1+n\sigma_W^2/\sigma^2)}} . \label{eq:RB_lb_Gauss}
\end{align}
\end{corollary}
\begin{IEEEproof}
Appendix~\ref{appd:GMean_GPrior_BernBias}.
\end{IEEEproof}

\noindent Note that the MMAE in Example~\ref{ex:GaussMean_GaussPrior} is upper-bounded by 
\begin{align}
R_{\rm B} \le \sqrt{\frac{\sigma_W^2}{1+n\sigma_W^2/\sigma^2}} 
\end{align}
which is achieved by $\wh W = \E[W|X^n]$.
Thus the non-asymptotic lower bound on the Bayes risk in \eqref{eq:RB_lb_Gauss} captures the correct dependence on $n$, and is off from the true Bayes risk by a constant factor. 
If we apply the unconditional lower bound \eqref{eq:RBlb_mi} to Example~\ref{ex:GaussMean_GaussPrior}, we can only get an asymptotic lower bound
\begin{align}
R_{\rm B} \gtrsim \dfrac{1}{4\log\left(1+{n\sigma_W^2}/{\sigma^2}\right)}\sqrt{\dfrac{\pi\sigma_W^2}{1+n\sigma_W^2/\sigma^2}}  \qquad\text{as $n\rightarrow \infty$} 
\end{align}
which differs from the upper bound by a logarithmic factor in $n$.
This example shows that the conditional lower bound \eqref{eq:RBlb_condMI} can provide tighter results than its unconditional counterpart \eqref{eq:RBlb_mi}.

\begin{example}[Estimating Bernoulli bias with uniform prior]\label{ex:WU[01]BernW}
Consider the example where the parameter $W \sim U[0,1]$, the samples $X_i \sim {\rm Bern}(w)$ conditional on $W=w$ for $i=1,\ldots,n$, and $\ell(w,\wh w) = |w-\wh w|$.
\end{example}
\begin{corollary}\label{co:WU[01]BernW}
In Example~\ref{ex:WU[01]BernW}, the Bayes risk is lower bounded by
\begin{align}
R_{\rm B} &\gtrsim \frac{1}{16\sqrt{2\pi n}}  \qquad\text{as $n \rightarrow \infty$} . \label{eq:RB_lb_Bern}
\end{align}
\end{corollary}
\begin{IEEEproof}
Appendix~\ref{appd:GMean_GPrior_BernBias}.
\end{IEEEproof}
\noindent Note that the MMAE in Example~\ref{ex:WU[01]BernW} is upper bounded by
\begin{align}
R_{\rm B} \le \frac{1}{\sqrt{6n}} 
\end{align}
achieved by the sample mean estimator $\wh{W} = \frac{1}{n}\sum^n_{i=1}X_i$. Thus, the lower bound in \eqref{eq:RB_lb_Bern} asymptotically captures the correct dependence on $n$, and is off from the true Bayes risk by a constant factor. 

\subsection{Lower bounds based on information density and small ball probability}\label{sec:RB_lb_cent_id}
For a joint distribution $P_{U,W,X}$ on $\sU\times\sW \times \sX$, define the conditional information density as
\begin{align}
i(w;x|u) 
= \log\frac{{\rm d}P_{W|U=u,X=x}}{{\rm d}P_{W|U=u}}(w) .
\end{align}
We give a lower bound on the probability of excess distortion in terms of conditional information density and conditional small ball probability:
\begin{lemma}\label{lm:excess_lb_iden}
For any estimate $\wh W$ of $W$ based on the sample $X$, any $\rho, \gamma>0$, and any auxiliary random variable $U$,
\begin{align}\label{eq:excessP_sbp_np}
\PP[\ell(W,\wh W) \ge \rho]  \ge  &\PP[i(W;X|U) < \log\gamma] - \gamma \E[{\mathcal L}_{W|U}(U,\rho)] + \nonumber \\
&\quad \gamma \inf_{u,w,x}\frac{{\rm d}P_{W|U=u}}{{\rm d}P_{W|U=u,X=x}}(w) \, \PP[i(W;X|U) \ge \log\gamma]   .
\end{align}
\end{lemma}
\begin{IEEEproof}The proof, inspired by the metaconverse technique from \cite{PPV10}, is given in Appendix~\ref{appd:RB_lb_NP}.
\end{IEEEproof}
Our second Bayes risk lower bound for centralized estimation is a consequence of Lemma~\ref{lm:excess_lb_iden}:
\begin{theorem}\label{th:RB_sbp_NP}
The Bayes risk for estimating the parameter $W$ based on the sample $X$ with respect to the distortion function $\ell$ satisfies
\begin{align}\label{eq:RB_sbp_NP_cond}
R_{\rm B}
&\ge \sup_{P_{U|W,X}} \sup_{\rho,\gamma>0} \rho \big(\PP[i(W;X|U) < \log\gamma] - \gamma \E[{\mathcal L}_{W|U}(U,\rho)]\big) .
\end{align}
In particular,
\begin{align}\label{eq:RB_sbp_NP_uncond}
R_{\rm B}
&\ge \sup_{\rho,\gamma>0} \rho \big(\PP[i(W;X) < \log\gamma] - \gamma {\mathcal L}_W(\rho)\big) .
\end{align}
\end{theorem}
\begin{IEEEproof}
With the aid of Markov's inequality, \eqref{eq:excessP_sbp_np} leads to the inequality
\begin{align}\label{eq:RB_sbp_NP_stronger}
R_{\rm B}
\ge \sup_{P_{U|W,X}} \sup_{\rho,\gamma>0} \rho \bigg(&\PP[i(W;X|U) < \log\gamma] - \gamma \E[{\mathcal L}_{W|U}(U,\rho)] + \nonumber \\
&\gamma \inf_{u,w,x}\frac{{\rm d}P_{W|U=u}}{{\rm d}P_{W|U=u,X=x}}(w) \PP[i(W;X|U) \ge \log\gamma] \bigg) .
\end{align}
The lower bound in \eqref{eq:RB_sbp_NP_cond} follows by replacing $ \inf_{u,w,x}\frac{{\rm d}P_{W|U=u}}{{\rm d}P_{W|U=u,X=x}}(w)$ with $0$.
\end{IEEEproof}

We give a high-dimensional example to illustrate the usefulness of Theorem~\ref{th:RB_sbp_NP}:

\begin{example}[Estimating $d$-dimensional Gaussian mean with uniform prior on $d$-ball]\label{ex:GaussMean_UPrior}
Consider the case where the parameter $W\in\R^d$ is distributed uniformly on the ball $\sW = \{ w \in \R^d: \|w \|_2 \le a \}$, the samples are $X_i = W + Z_i$ with $Z_i \sim N(0,\sigma^2 {\mathbf I}_d)$ independent of $W$ for $i=1,\ldots,n$, and $\ell(w,\wh w) = \|w-\wh w\|_2$.
\end{example}
\begin{corollary}\label{co:GaussMean_UPrior}
In Example~\ref{ex:GaussMean_UPrior}, for any $a>0$, $\sigma^2 >0$, and $d\ge 1$, the Bayes risk is lower bounded by
\begin{align}
R_{\rm B}&\gtrsim \frac{1}{20} \sqrt{\frac{2\pi\sigma^2 d}{n}} \qquad\text{as $n\rightarrow\infty$} \label{eq:RB_lb_UGauss_WU} .
\end{align}
\end{corollary}
\begin{IEEEproof}
Appendix~\ref{appd:dGMean_UPrior}.
\end{IEEEproof}
\noindent Note that the Bayes risk in Example~\ref{ex:GaussMean_UPrior} is upper bounded by
\begin{align}
R_{\rm B}
\le \sqrt{\frac{\sigma^2 d}{n}} 
\end{align}
achieved by the sample mean estimator $\widehat{W} = \frac{1}{n}\sum^n_{i=1}X_i$. Thus, the lower bound in \eqref{eq:RB_lb_UGauss_WU} captures the correct dependence on $n$ (asymptotically) and $d$ (non-asymptotically), and is off from the true Bayes risk by a constant factor.
Moreover, by squaring \eqref{eq:RB_lb_UGauss_WU}, we get a lower bound on the MMSE that also captures the correct dependence on $n$ and $d$.

\subsection{Generalizations of Fano's inequality}\label{sec:RB_lb_cent_Fano}
The lower bounds on the probability of excess distortion in Lemmas~\ref{lm:excess_lb_csbp} and \ref{lm:excess_lb_iden} can be viewed as generalizations of Fano's inequality. 

When $W$ takes  values on $\{1,\ldots,M\}$ and $\ell(w,\wh w) = \I\{w \neq \wh w\}$, setting $\rho=1$ in \eqref{eq:1-pt_lb} without conditioning on $U$ recovers the following generalization of Fano's inequality due to Han and Verd\'u \cite{Han_Ver_Fano}:
\begin{align}\label{eq:Pe_lb_mi}
\PP[\wh W \neq W] \ge 1 - \frac{I(W;X)+1}{\log(1/\max_{w\in [M]}P_W(w))} .
\end{align}
Similarly, setting $\rho=1$ in \eqref{eq:excessP_sbp_np} without conditioning on $U$, we get
\begin{align}\label{eq:Pe_lb_id}
\PP[\wh W \neq W] 
\ge \sup_{\gamma>0} \, \PP[i(W;X) < \log\gamma] - \gamma {\mathcal L}_{W}(\rho) + 
\gamma \inf_{w,x}\frac{{\rm d}P_{W}}{{\rm d}P_{W|X=x}}(w) \PP[i(W;X) \ge \log\gamma] .
\end{align}
When $W$ is uniformly distributed on $\{1,\ldots,M\}$, \eqref{eq:Pe_lb_mi} reduces to the usual Fano's inequality 
\begin{align}
\PP[\wh W \neq W] \ge 1 - \frac{I(W;X)+1}{\log M} ,
\end{align}
while \eqref{eq:Pe_lb_id} reduces to the Poor--Verd\'u bound \cite{Poo_Ver_Pe}
\begin{align}
\PP[\wh W \neq W] 
\ge \sup_{\gamma>0} \, \left(1-\frac{\gamma}{M}\right)\PP[i(W;X) < \log\gamma].
\end{align}

When $W$ is continuous, Eqs.~\eqref{eq:1-pt_lb} and \eqref{eq:excessP_sbp_np}  provide continuum generalizations of Fano's inequality. 
For example, when $\sW \subset \R^d$ and $\ell(w,\wh w) = \|w - \wh w\|_2$, \eqref{eq:1-pt_lb} leads to
\begin{align}
\PP\big[\|\wh W - W\|_2 \ge \rho\big] \ge 1 - \frac{I(W;X)+1}{\log (1/\sup_{w\in\sW}\PP[\|W-w\|_2 < \rho])} 
\end{align}
which is also obtained by Chen et al.~\cite{Chen_Bayes}, and generalizes the result of Duchi and Wainwright \cite{DucWai_Fano}.
Similarly, \eqref{eq:excessP_sbp_np} leads to
\begin{align}
\PP[\|\wh W - W\|_2 \ge \rho]  &\ge  \sup_{\gamma>0} \left( \PP[i(W;X) < \log\gamma] - \gamma \sup_{w\in\sW}\PP[\|W-w\|_2 < \rho] \right).
\end{align}

\subsection{Lower bounds based on mutual information and differential entropy}\label{sec:RB_lb_cent_diff}
For the problem of estimating a real-valued parameter $W$ with respect to the quadratic distortion $\ell(w,\wh{w}) = |w-\wh{w}|^2$, it can be shown that (see, e.g.,~\cite[Lemma 5]{AXMR_dist_comp}), if $\E(W-\wh W)^2 \le \alpha$, then
\begin{align}
I(W;\wh W|U) \ge h(W|U) - \frac{1}{2}\log({2\pi e \alpha}) .
\end{align}
Upper-bounding $I(W;\wh W|U)$ by $I(W;X|U)$, we obtain a lower bound on the MMSE
\begin{align}
\inf_{\psi} \E (W-\wh W)^2 \ge \sup_{P_{U|W,X}} \frac{1}{2\pi e} 2^{-2(I(W;X|U) - h(W|U))} .
\end{align}
More generally, for the problem of estimating a parameter $W$ taking values in $\R^d$,  the Shannon lower bound on the rate-distortion function (see, e.g.,~\cite[Chap.\ 4.8]{Gray_source_coding}) can be used to show that, if $\E \|W-\wh W\|^r \le \alpha$ with an arbitrary norm $\|\cdot\|$ in $\R^d$ and an arbitrary $r\ge 1$, then
\begin{align}\label{eq:RD_lb_diff_vol}
I(W;\wh W) \ge h(W) - \log \left(V_d \Big(\frac{\alpha re}{d}\Big)^{d/r} \Gamma\Big(1 + \frac{d}{r}\Big) \right) ,
\end{align}
where $V_d$ is the volume of the unit ball in $(\R^d, \| \cdot \|)$ and $\Gamma(\cdot)$ is the gamma function. For example, this method can be used to recover the lower bounds of Seidler \cite{Seidler71} for the problem of estimating a parameter in $\R^d$ with respect to squared weighted $\ell_2$ norms, and gives tight lower bounds on the Bayes risk and the minimax risk in high-dimensional estimation problems \cite[Lec. 13]{Wu_lec_stat}.  A simple extension of \eqref{eq:RD_lb_diff_vol} via an auxiliary random variable $U$ gives
\begin{align}\label{eq:RD_lb_cond_diff_vol}
I(W;\wh W|U) \ge h(W|U) - \log \left(V_d \Big(\frac{\alpha re}{d}\Big)^{d/r} \Gamma\Big(1 + \frac{d}{r}\Big) \right) .
\end{align}
As a consequence, we obtain a lower bound on the Bayes risk in terms of conditional mutual information and conditional differential entropy:
\begin{theorem}\label{th:RB_lb_mi_diff} 
For an arbitrary norm $\|\cdot\|$ in $\R^d$ and any $r\ge 1$, the Bayes risk for estimating the parameter $W \in \R^d$ based on the sample $X$ with respect to the distortion function $\ell(w,\wh w) = \|w-\wh w\|^r$ satisfies
\begin{align}\label{eq:RB_lb_mi_diff}
R_{\rm B}
&\ge \sup_{P_{U|W,X}} \frac{d}{re} \left(V_d \Gamma\Big(1+\frac{d}{r}\Big)\right)^{-{r}/{d}} 2^{-(I(W;X|U) - h(W|U))r/d} .
\end{align}
In particular, for estimating a real-valued $W$ with respect to $\ell(w,\wh w) = |w-\wh w|$,
\begin{align}
R_{\rm B} \ge \sup_{P_{U|W,X}} \frac{1}{2 e} 2^{-(I(W;X|U) - h(W|U))} .
\end{align}
\end{theorem}
The advantage of Theorem~\ref{th:RB_lb_mi_diff} is that its unconditional version can yield tighter Bayes risk lower bounds than the unconditional version of Theorem~\ref{th:RB_lb_mi}.
For example, consider the case where $W$ is uniformly distributed on $[0,1]$, and is estimated based on $X$ with respect to the absolute distortion. 
Setting $g(\rho) = 2\rho$ in Remark~\ref{rmk:th:RB_lb_mi} and optimizing $s$ in \eqref{eq:RB_sup_s}, the unconditional version of Theorem~\ref{th:RB_lb_mi} yields an asymptotic lower bound
\begin{align}
R_{\rm B} \gtrsim \frac{1}{8I(W;X)}2^{-I(W;X)} \qquad\text{as $I(W;X)\rightarrow \infty$.} \label{eq:RB_cont_asymp}
\end{align}
By contrast, the unconditional version of Theorem~\ref{th:RB_lb_mi_diff} yields a tighter and non-asymptotic lower bound
\begin{align}\label{eq:RB_WU01_abs}
R_{\rm B} \ge \frac{1}{2 e} 2^{-I(W;X)} .
\end{align}

\section{Mutual information contraction via SDPI}\label{sec:SDPI_mi}
While the results in Section~\ref{sec:RB_lb_central} all apply to general estimation problems, either centralized or decentralized, the results in terms of mutual information (Theorems~\ref{th:RB_lb_mi} and \ref{th:RB_lb_mi_diff}) are particularly amenable to tightening in the context of the decentralized estimation.
For example, Theorem~\ref{th:RB_lb_mi} reveals two sources of the difficulty of estimating $W$: the spread of the prior distribution $P_W$ or its conditional counterpart $P_{W|U}$, captured by $\cL_{\mathcal W}$ or $\cL_{W|U}$, and the amount of information about $W$ contained in the sample $X$, captured by $I(W;X)$ or $I(W;X|U)$. 
When an estimator does not have direct access to $X$, but can only receive information about it from one or more local processors, the amount of information about $W$ contained in the estimator's indirect observations will contract relative to $I(W;X)$ or $I(W;X|U)$. The contraction is caused by the communication constraints between the local processors and the estimator, such as finite precision of analog-to-digital conversion, storage limitations of intermediate results, limited transmission blocklength, channel noise, etc. 

We will quantify this contraction of mutual information through strong data processing inequalities, or SDPI's, for the relative entropy (see \cite{MR_SDPI} and references therein).
Given a stochastic kernel (or channel) $K$ with input alphabet $\sX$ and output alphabet $\sY$ and a reference input distribution $\mu$ on $\sX$, we say that $K$ satisfies an SDPI at $\mu$ with constant $c\in[0,1)$ if $D(\nu K \| \mu K) \le c D(\nu \| \mu)$ for any other input distribution $\nu$ on $\sX$. Here, $\mu K$ denotes the marginal distribution of the channel output when the input has distribution $\mu$. The SDPI constants of $K$ are defined by
\begin{align*}
\eta(\mu,K) \deq \sup_{\nu:\, \nu\neq\mu}\frac{D(\nu K \| \mu K)}{D(\nu \| \mu)}, \qquad
\eta(K) \deq \sup_{\mu}\eta(\mu,K).
\end{align*}
It is shown in \cite{VA_SDPI} that the SDPI constants are also the maximum contraction ratios of mutual information in a Markov chain:
for a Markov chain $W - X - Y$, 
\begin{align}\label{eq:MI_SDPI_Px}
\sup_{P_{W|X}} \frac{I(W;Y)}{I(W;X)} = \eta(P_X,P_{Y|X}) 
\end{align}
if the joint distribution $P_{X,Y}$ is fixed, and
\begin{align}\label{eq:MI_SDPI}
\sup_{P_{W,X}} \frac{I(W;Y)}{I(W;X)} = \eta(P_{Y|X}) 
\end{align}
if only the channel $P_{Y|X}$ is fixed. 
This fact leads to the following SDPI's for mutual information:
\begin{align}\label{eq:MI_SDPI_WXY}
I(W;Y) \le I(W;X)\eta(P_X,P_{Y|X}) \le I(W;X)\eta(P_{Y|X}) .
\end{align}
It is generally hard to compute the SDPI constant for an arbitrary pair of $\mu$ and $K$, except for some special cases:
\begin{itemize}
	\item For the binary symmetric channel, $\eta({\rm Bern}(\frac{1}{2}),{\rm BSC}(\eps)) = \eta({\rm BSC}(\eps)) = (1-2\eps)^2$ \cite{Ahlswede_Gacs_hypercont}.
	\item For the binary erasure channel, $\eta({\rm Bern}(\frac{1}{2}),{\rm BEC}(\eps)) = \eta({\rm BEC}(\eps)) = 1-\eps$. 
	\item If $X$ and $Y$ are jointly Gaussian with correlation coefficient $\rho_{X,Y}$, then \cite{CovErk98}
\begin{align}\label{eq:SDPI_GaussXY}
\eta(P_X,P_{Y|X}) = \rho_{X,Y}^2 .
\end{align}
\end{itemize}
In the remainder of this section, we collect a few upper bounds and properties of the SDPI constants, which will be used in the sequel. The first upper bound is due to Cohen et al.~\cite{Cohen_SDPI}:
\begin{lemma}\label{lm:Dobru_coef}
Define the {\em Dobrushin contraction coefficient} of a stochastic kernel $P_{X|W}$ by
\begin{align}
\vartheta(P_{X|W}) = \max_{w,w'}\|P_{X|W=w}-P_{X|W=w'}\|_{\rm TV}.
\end{align} 
Then 
\begin{align}
\eta(P_{X|W}) \le \vartheta(P_{X|W}) .
\end{align} 
\end{lemma}
\noindent The next upper bound is proved in \cite[Remark~3.2]{MR_SDPI} for arbitrary $f$-divergences:
\begin{lemma}\label{lm:Doeblin}
Suppose there exist a constant $\alpha\in(0,1]$ and a distribution $Q_X$, such that\footnote{In Markov chain theory, this is known as a Doeblin minorization condition.}
\begin{align}
\frac{{\rm d}P_{X|W=w}}{{\rm d}Q_X}(x) \ge \alpha \qquad\text{for all $x\in\sX$ and $w\in\sW$}. 
\end{align}
Then
\begin{align}
\eta(P_{X|W}) \le 1-\alpha .
\end{align} 
\end{lemma}
\noindent Lemma~\ref{lm:Doeblin} leads to the following property:
\begin{lemma}\label{lm:back_ch}
For a joint distribution $P_{W,X}$, suppose there is a constant $\alpha\in(0,1]$ such that the forward channel $P_{X|W}$ satisfies
\begin{align}\label{eq:Doeblin_forbak_cond}
\frac{{\rm d}P_{X|W=w}}{{\rm d}P_{X|W=w'}}(x) \ge {\alpha} \qquad\text{for all $x\in\sX$ and $w,w'\in\sW$.}
\end{align}
Then the SDPI constants of the forward channel $P_{X|W}$ and the backward channel $P_{W|X}$ satisfy 
\begin{align}
\eta(P_{X|W}) \le 1-\alpha \quad\text{and}\quad \eta(P_{W|X}) \le 1-\alpha .
\end{align}
\end{lemma}
\begin{IEEEproof} 
To prove the claim for the forward channel, pick any $w'\in\sW$ and let $Q_X = P_{X|W=w'}$. Then the condition in Lemma~\ref{lm:Doeblin} is satisfied with this $Q_X$.
To prove the claim for the backward channel, consider any $x\in\sX$ and $w\in\sW$. Then
\begin{align}
\frac{{\rm d}P_{W|X=x}}{{\rm d}P_{W}}(w)
&= \frac{{\rm d} P_{X|W=w}}{{\rm d} \int P_{X|W=w'}  P_W({\rm d}w')}(x) \\
&= \frac{1}{\int \frac{{\rm d} P_{X|W=w'}}{{\rm d} P_{X|W=w}}(x)  P_W({\rm d}w')} \\
&\ge \frac{1}{\frac{1}{\alpha}\int P_W({\rm d}w')} \label{eq:Doeblin_forbak1}\\
&= \alpha ,
\end{align}
where \eqref{eq:Doeblin_forbak1} uses the fact that $\frac{{\rm d} P_{X|W=w'}}{{\rm d} P_{X|W=w}}(x) \le 1/\alpha$, due to the assumption in \eqref{eq:Doeblin_forbak_cond}.
Using Lemma~\ref{lm:Doeblin} with $Q_W = P_W$, we get the result. 
\end{IEEEproof}

In decentralized estimation, we will encounter the SDPI constant $\eta(P_{X^{n}},P_{W|X^{n}})$. The following lemma gives an upper bound for this SDPI constant, which is often easier to compute:
\begin{lemma}\label{lm:SDPI_ub_ss}
If $W - Z - X^n$ form a Markov chain, then
\begin{align}
\eta(P_{X^{n}},P_{W|X^{n}}) \le \eta(P_{Z},P_{W|Z}) .
\end{align}
In particular, $Z$ can be any sufficient statistic of $X^n$ for estimating $W$.
\end{lemma}
\begin{IEEEproof}
It suffices to show that for any $Y$ such that $W - X^n - Y$ form a Markov chain,
\begin{align}\label{eq:SDPI_ss_sc}
I(W;Y) &\le \eta(P_{Z},P_{W|Z}) I(X^n;Y) .
\end{align}
Indeed, by the definition of $\eta(P_{Z},P_{W|Z})$ and the fact that $W - Z - X^n - Y$ form a Markov chain,
\begin{align}
I(W;Y) &\le \eta(P_{Z},P_{W|Z}) I(Z;Y) \\
&\le \eta(P_{Z},P_{W|Z}) I(X^n;Y) ,
\end{align}
which proves \eqref{eq:SDPI_ss_sc} and the lemma.
\end{IEEEproof}

For product input distributions and product channels, the SDPI constant tensorizes \cite{VA_SDPI} (see \cite{MR_SDPI} for a more general result for other $f$-divergences):
\begin{lemma}\label{lm:SDPI_tensor}
For distributions $\mu_1,\ldots,\mu_n$ on $\sX$ and channels $K_1,
\ldots,K_n$ with input alphabet $\sX$,
\begin{align*}
\eta( \mu_1 \otimes \ldots \otimes \mu_n , K_1 \otimes \ldots \otimes K_n) = \max_{1\le i\le n} \eta(\mu_i, K_i) .
\end{align*}
\end{lemma}
\noindent Finally, the following lemma due to Polyanskiy and Wu \cite{PolWu_ES}
gives an SDPI for multiple uses of a channel:
\begin{lemma}\label{lm:noisy_ch_SDPI}
Consider sending a message $Y$ through $T$ uses of a memoryless channel $P_{V|U}$ with feedback, where $U_t = \varphi(Y,V^{t-1},t)$ with some encoder $\varphi$ for $t=1,\ldots,T$.
Then for any random variable $W$ such that $W - Y - U^T, V^T$ form a Markov chain,
\begin{align}\label{eq:SDPI_DMCfb}
I(W;V^T) \le I(W;Y) \big(1 - (1-\eta(P_{V|U}))^T\big) .
\end{align}
In particular, the result holds when the channel is used $T$ times without feedback.
\end{lemma}
\begin{IEEEproof}
Let $\eta = \eta(P_{V|U})$. Then
\begin{align}
I(W;V^T) &= I(W;V^{T-1}) + I(W;V_T|V^{T-1}) \\
&\le I(W;V^{T-1}) + \eta I(W;U_T|V^{T-1}) \label{eq:IWVtUt_cond} \\
&= (1-\eta)I(W;V^{T-1}) + \eta I(W;V^{T-1},U_T) \\
&\le (1-\eta)I(W;V^{T-1}) + \eta I(W;Y) \label{eq:IWVU_IWY} ,
\end{align}
where \eqref{eq:IWVtUt_cond} follows from the Markov chain $W,V^{T-1} - U_T - V_T$ and a conditional version of SDPI \cite[Lemma~1]{AXMR_dist_comp};  \eqref{eq:IWVU_IWY} follows from the Markov chain $W - Y - V^{T-1},U_T$. 
Unrolling the above recursive upper bound on $I(W;V^T)$ and noting that $I(W;V_1) \le \eta I(W;Y)$, we get (\ref{eq:SDPI_DMCfb}).
\end{IEEEproof}
Using the same proof technique, it can be shown that \cite[Lemma~2]{AXMR_dist_comp} for the $T$\,th product of a channel $P_{V|U}$,
\begin{align}\label{eq:SDPI_TthPV|U}
\eta(P_{V|U}^{\otimes T}) \le 1 - (1-\eta(P_{V|U}))^T .
\end{align}

\section{Decentralized estimation: single processor setup}\label{sec:single}

We start the discussion of  decentralized estimation with the single-processor setup.
Consider the following decentralized estimation problem with one local processor, shown schematically in Fig.~\ref{fg:model_single_bT}:

\begin{figure}[h!]
\centering
  \includegraphics[scale = 1.15]{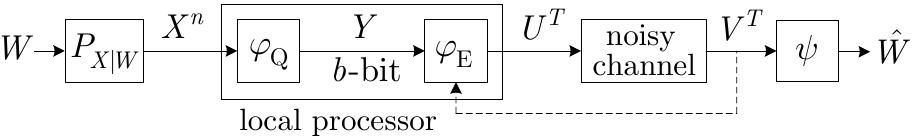}
  \caption{Model of decentralized estimation (single processor).}
  \label{fg:model_single_bT}
\end{figure}

\begin{itemize}
\item
$W$ is an unknown parameter (discrete or continuous, scalar or vector) with prior distribution $P_W$.
\item
Conditional on $W=w$, $n$ samples $X^n = (X_1,\ldots,X_n)$ are independently drawn from the distribution $P_{X|W=w}$.
\item
The local processor observes $X^n$ and maps it to a $b$-bit message $Y=\varphi_{\rm Q}(X^n)$.
\item
The encoder maps $Y$ to a codeword $U^T = \varphi_{\rm E}(Y)$ with blocklength $T$, and transmits $U^T$ over a discrete memoryless channel (DMC) $P_{V|U}$. 
We allow the possibility of feedback from the estimator to the processor, in which case $U_t = \varphi_{\rm E}(Y,V^{t-1},t)$, $t = 1,\ldots,T$.
\item
The estimator computes $\wh W = \psi(V^T)$ as an estimate of $W$, based on the received codeword $V^T$.
\end{itemize}
The Bayes risk in the single processor setup is defined as 
\begin{align}
R_{\rm B} = \inf_{\varphi_{\rm Q},\varphi_{\rm E},\psi} \E\big[\ell(W,\psi(V^T))\big] , 
\end{align}
which depends on the problem specification including $P_{W,X}$, $\ell$, $n$, $b$, $T$, and $P_{V|U}$.
We can use the unconditional versions of Theorems~\ref{th:RB_lb_mi} and \ref{th:RB_lb_mi_diff} to obtain lower bounds for $R_{\rm B}$, by replacing $I(W;X)$ with $I(W;V^T)$. To reveal the dependence of $R_{\rm B}$ on various problem specifications, we need an upper bound on $I(W;V^T)$ which is independent of $\varphi_{\rm Q}$ and $\varphi_{\rm E}$:
\begin{theorem}\label{th:gen_single}
In decentralized estimation with a single processor, for any choice of $\varphi_{\rm Q}$ and $\varphi_{\rm E}$,
\begin{align}\label{eq:miub_gen_single}
I(W;V^T) & \le  \min\Big\{I(W;X^n)\eta_T, \,\, \eta(P_{X^n},P_{W|X^n}) \left(H(X^n) \wedge b \right) \eta_T, \,\,
\eta(P_{X^n},P_{W|X^n}) CT \Big\}
\end{align}
where $C$ is the Shannon capacity of the channel $P_{V|U}$, and 
\begin{align}\label{eq:eta_chT}
\eta_T = 
\begin{cases}
1-(1-\eta(P_{V|U}))^T &\quad \text{with feedback} \\
\eta(P_{V|U}^{\otimes T}) &\quad \text{without feedback}
\end{cases} .
\end{align}
\end{theorem}
\begin{IEEEproof} 
When the channel is used with feedback, the problem setup gives rise to the Markov chain $W - X^n - Y - U^T, V^T$. 
With $\eta_T = 1-(1-\eta(P_{V|U}))^T$, as a consequence of Lemma~\ref{lm:noisy_ch_SDPI}, we have
\begin{align}
I(W;V^T) &\le I(W;Y)\eta_T 
\le I(W;X^n)\eta_T.  \label{eq:MI_ub_WYWU}
\end{align}
Alternatively, 
\begin{align}
I(W;V^T) &\le I(W;Y)\eta_T  \\
&\le \eta(P_{X^n},P_{W|X^n}) I(X^n;Y) \eta_T \label{eq:I_WY_1} \\
&\le \eta(P_{X^n},P_{W|X^n}) (H(X^n) \wedge b) \eta_T \label{eq:mi_ub_etabeta} 
\end{align}
where \eqref{eq:I_WY_1} is from the SDPI in (\ref{eq:MI_SDPI_WXY}); \eqref{eq:mi_ub_etabeta}
is because $I(X^n;Y)\le \min\{H(X^n),H(Y)\}$ and $Y \in [2^b]$.
Lastly, from the SDPI and following the proof that feedback does not increase the capacity of a discrete memoryless channel \cite{Cover_book},
\begin{align*}
I(W;V^T) &\le \eta(P_{X^n},P_{W|X^n}) I(Y;V^T) \\
&\le \eta(P_{X^n},P_{W|X^n}) CT . 
\end{align*}
We complete the proof for the case with feedback by taking the minimum of the three resulting estimates to get the tightest bound on $I(W;V^T)$.

When the channel is used without feedback, we have the Markov chain $W - X^n - Y - U^T - V^T$.
In this case, \eqref{eq:MI_ub_WYWU} holds with $\eta_T = \eta(P_{V|U}^{\otimes T})$ as a conequence of the SDPI. 
The rest of the proof for this case is the same as the case with feedback.
\end{IEEEproof}

Note that, with the ordinary data processing equality, we can only get the upper bound
\begin{align}\label{eq:miub_gen_single_ODPI}
I(W;V^T) & \le  \min\Big\{I(W;X^n), \, H(X^n) \wedge b , \,  CT \Big\} ,
\end{align}
where the first term reflects the statistical constraint due to the finite number of samples, the second term reflects the communication constraint due to the quantization, and the third term reflects the communication constraint due to the noisy channel.
All of these terms are tightened in \eqref{eq:miub_gen_single} via the multiplication by various contraction coefficients. 
Thus, using the SDPI, we can tighten the results of Theorems~\ref{th:RB_lb_mi} and \ref{th:RB_lb_mi_diff} in the setting of decentralized estimation by quantifying the communication constraint, and by coupling the statistical constraint and the communication constraint together.

Next we study a few examples of this problem setup to illustrate the effectiveness of using Theorem~\ref{th:gen_single} to derive lower bounds on the Bayes risk.

\subsection{Transmitting a bit over a BSC}
\begin{example}\label{ex:1bit_1sender}
Consider the case where the parameter takes values $0$ and $1$ with equal probabilities, the local processor directly observes $W$, and communicates the value of $W$ to the estimator through $T$ uses of the channel ${\rm BSC}(\eps)$. Formally, $W$ is ${\rm Bern}(\frac{1}{2})$, $W=X^n=Y$, and $P_{V|U}={\rm BSC}(\eps)$. The Bayes risk is defined as $R_{\rm B} = \inf_{\varphi_{\rm E},\psi}\PP[W \neq \wh W]$.
\end{example}
In this simple example, there is no statistical constraint since $W$ can be directly observed by the local processor, while the communication constraint is imposed by the $T$ uses of a BSC.
Using Theorem~\ref{th:gen_single}, we can derive lower bounds on $R_{\rm B}$ and obtain upper bounds on the error exponent when the channel is used with or without feedback:
\begin{corollary}\label{co:1bit_1sender}
In Example~\ref{ex:1bit_1sender}, if the channel is used without feedback, then
\begin{align}\label{eq:BSC1b_plb}
R_{\rm B} \ge h_2^{-1}\left(\frac{1}{\sqrt{2T}} (4\eps(1-\eps))^{\frac{T}{2}}\right) ,
\end{align}
and
\begin{align}\label{eq:BSC1b_expub}
\limsup_{T\rightarrow\infty} - \frac{1}{T} \log R_{\rm B} \le \frac{1}{2}\log\frac{1}{4\eps(1-\eps)} .
\end{align}
If the channel is used with feedback, then
\begin{align}\label{eq:BSC1b_plb_fb}
R_{\rm B} \ge h_2^{-1}\left((4\eps(1-\eps))^{T}\right) ,
\end{align}
and
\begin{align}\label{eq:BSC1b_expub_fb}
\limsup_{T\rightarrow\infty} - \frac{1}{T} \log R_{\rm B} \le \log\frac{1}{4\eps(1-\eps)} .
\end{align}
\end{corollary}
\begin{IEEEproof}
Choose the $\varphi_{\rm E}$ and $\psi$ that attain $R_{\rm B}$.
In this case, we can bypass Theorem~\ref{th:RB_lb_mi} by using the binary-alphabet version of Fano's inequality:
\begin{align}\label{eq:plb_1-h}
1 - h_2(\PP[\wh W\neq W]) \le I(W;V^T) .
\end{align}
If the channel is used without feedback, it follows from Theorem~\ref{th:gen_single} and Lemma~\ref{lm:Dobru_coef} that
\begin{align}
I(W;V^T) &\le I(W;X^n) \eta\big({\rm BSC}(\eps)^{\otimes T}\big)
\le \vartheta\big({\rm BSC}(\eps)^{\otimes T}\big) 
\le 1 - \frac{1}{\sqrt{2T}} (4\eps(1-\eps))^{T/2} , \label{eq:BSC1b_miub}
\end{align}
where the upper bound on $\vartheta\big({\rm BSC}(\eps)^{\otimes T}\big)$ is evaluated in \cite{PolWu_ES}.
Combining \eqref{eq:plb_1-h} and \eqref{eq:BSC1b_miub}, and using the fact that \cite[Theorem 2.2]{Calabro_thesis}
\begin{align}\label{eq:lb_h2inv}
h_2^{-1}(x) \ge \frac{x}{2\log({6}/{x})} \qquad \text{for $x\in [0,1]$},
\end{align}
we obtain \eqref{eq:BSC1b_plb} and \eqref{eq:BSC1b_expub}.

If the channel is used with feedback, Theorem~\ref{th:gen_single} gives
\begin{align}\label{eq:BSC1b_miub_fb}
I(W;V^T) &\le I(W;X^n) \big(1 - (1-\eta({\rm BSC}(\eps))^T\big) 
\le 1 - (4\eps(1-\eps))^{T} 
\end{align}
where we used the fact that $\eta({\rm BSC}(\eps)) = (1-2\eps)^2$.
Combining \eqref{eq:plb_1-h}, \eqref{eq:lb_h2inv} and \eqref{eq:BSC1b_miub_fb}, we obtain \eqref{eq:BSC1b_plb_fb} and \eqref{eq:BSC1b_expub_fb}.
\end{IEEEproof}
Using the Chernoff bound, it can be shown that a blocklength-$T$ repetition code without feedback can achieve $\PP[\wh W\neq W] \le ({4\eps(1-\eps)})^{-{T}/{2}}$  \cite{Gallager_ITbook}. Thus, when the channel is used without feedback,
\begin{align}
\liminf_{T\rightarrow\infty} - \frac{1}{T} \log R_{\rm B} \ge \frac{1}{2}\log\frac{1}{4\eps(1-\eps)} ,
\end{align}
which matches the upper bound on the error exponent given by \eqref{eq:BSC1b_expub}.
Therefore, Theorem~\ref{th:gen_single} can effectively capture the communication constraint in this example.

\subsection{Estimating a discrete parameter}\label{sec:W{pm1}^dBSC}
\begin{example}\label{ex:W{pm1}^dBSC}
Consider the case where $W$ is uniformly distributed on $\{-1,1\}^d$.
The sample $X \in\{-1,1\}^{d}$ is generated conditionally on $W$ as follows.
For $j=1,\ldots,d$, given
$W_j = w_j$, the $j$th coordinate of of $X$, denoted by $X_j$, is independently drawn from the distribution $P_{X_j|W_j=w_j}(x_{j}) = ({1+x_{j}w_j\delta})/{2}$ for some  $\delta\in[0,1]$.
In other words, $P_{X_{j}|W_j}$ is $\rm{BSC}(\frac{1-\delta}{2})$. It follows that $X_j$ is uniformly distributed on $\{-1,1\}$, and $P_{W_j|X_{j}}$ is $\rm{BSC}(\frac{1-\delta}{2})$ as well. 
The communication channel $P_{V|U}$ is assumed to be an arbitrary DMC.
\end{example}
Theorem~\ref{th:gen_single} gives the following upper bound on $I(W;V^T)$ for this example:
\begin{corollary}\label{co:W_disc}
In Example~\ref{ex:W{pm1}^dBSC},
\begin{align}
I(W;V^T) &\le \min\Big\{d\big(1-h_2\big(\tfrac{1-\delta}{2}\big)\big)\eta_T, \,\,\delta^2 b\eta_T, \,\, \delta^2 CT\Big\}. \label{eq:W_disc_n=1}
\end{align}
\end{corollary}
\begin{IEEEproof}
Since $(W_1,X_1),\ldots,(W_d,X_d)$ are independent in this case, we can apply the tensorization property of the SDPI constant (Lemma~\ref{lm:SDPI_tensor}), which states that
\begin{align}
\eta(P_{X},P_{W|X}) = \max_{1\le j\le d}\eta(P_{X_j},P_{W_j|X_j}) . 
\end{align}
Due to the fact that $X_{j}$ is uniform on $\{-1, 1\}$ and $P_{W_j|X_{j}}=\rm{BSC}(\frac{1-\delta}{2})$, we have the exact SDPI constant 
\begin{align}
\eta(P_{X_j},P_{W_j|X_j}) = \delta^2 .
\end{align}
We also have $I(W;X) = d\big(1-h_2\big(\tfrac{1-\delta}{2}\big)\big)$.
The results then follow from Theorem~\ref{th:gen_single}.
\end{IEEEproof}
The same problem with noiseless communication channel was considered in \cite{Zhang_dist13}. The result in \cite[Lemma 3]{Zhang_dist13}, proved in a much more complicated way, shows that
\begin{align}
I(W;Y) &\le \frac{32\delta^2 (d \wedge b)}{(1-\delta)^4} \label{eq:W_disc_Duchi}
\end{align}
where the contraction coefficient is less than $1$ only when $\delta<0.133$.
By contrast, the contraction coefficient in (\ref{eq:W_disc_n=1}) never exceeds $1$.
Moreover, since $1-h_2\big(\tfrac{1-\delta}{2}\big)\le \delta^2$, the upper bound in (\ref{eq:W_disc_n=1}) is a considerable improvement on the one in (\ref{eq:W_disc_Duchi}) over all $\delta\in[0,1]$, especially for large $\delta$, under the same noiseless channel assumption ($\eta_T = 1$).
Corollary~\ref{co:W_disc} can also be used to derive lower bounds on the minimax risk of estimating the mean of an arbitrary probability distribution on the cube $[-1,1]^d$. We discuss this application in Section~\ref{sec:multi_ind_data}, in the multi-processor setup.

From another point of view, Example~\ref{ex:W{pm1}^dBSC} is essentially a problem of noisy lossy source coding \cite{DobTsy_noisylossy} of an i.i.d. ${\rm Bern}(\frac{1}{2})$ source of length $d$ observed through a ${\rm BSC}(\frac{1-\delta}{2})$, with an additional challenge of sending the quantized message over $T$ uses of another noisy channel.
Using Corollary~\ref{co:W_disc}, we can obtain lower bounds on the average bit error probability for estimating the source $W$ and on the quantization rate of the sample $X$:
\begin{corollary}\label{co:W_disc_RBrate}
In Example~\ref{ex:W{pm1}^dBSC}, let $\ell(w,\wh{w}) = \tfrac{1}{d}\sum^d_{j=1} {\bf 1}{\{w_j \neq \wh{w}_j\}}$. Then, 
\begin{align}\label{eq:WdXdBSC_plb} 
R_{\rm B} \ge h_2^{-1}\left(1-\frac{1}{d}\min\Big\{d\big(1-h_2\big(\tfrac{1-\delta}{2}\big)\big)\eta_T ,\,\, \delta^2 b\eta_T,\,\,\delta^2 CT\Big\}\right),
\end{align}
provided $b$, $d$, and $T$ are such that the argument of $h_2^{-1}(\cdot)$ lies in $[0,1]$.
Moreover, to achieve $R_{\rm B} \le p$, it is necessary that
\begin{align}\label{eq:WdXdBSC_blb} 
\frac{b}{d} \ge \frac{1-h_2(p)}{\delta^2 \eta_T} ,
\end{align}
where $\eta_T = 1 - (1-\eta(P_{V|U}))^T$.
\end{corollary}
\begin{IEEEproof}
Choose the $\varphi_{\rm Q}$, $\varphi_{\rm E}$ and $\psi$ that attain $R_{\rm B}$.
In this case, we can again bypass Theorem~\ref{th:RB_lb_mi} by using the following chain of inequalities to relate the average bit error probability with $I(W;V^T)$:
\begin{align}
1 - h_2(R_{\rm B}) &= d_2(R_{\rm B}\| \tfrac{1}{2}) \\
&\le \frac{1}{d}\sum_{j=1}^d d_2\big(\PP[W_j \neq \wh W_j] \| \tfrac{1}{2}\big) \label{eq:WdXdBSC_1} \\
&\le \frac{1}{d}\sum_{j=1}^d I(W_j;\wh W_j) \label{eq:WdXdBSC_2} \\
&\le \frac{1}{d}I(W;V^T) \label{eq:WdXdBSC_3} \\
&\le \frac{1}{d}\min\Big\{d\big(1-h_2\big(\tfrac{1-\delta}{2}\big)\big)\eta_T ,\, \delta^2 b\eta_T,\, \delta^2 CT\Big\} \label{eq:WdXdBSC_4}
\end{align}
where \eqref{eq:WdXdBSC_1} uses the fact that $R_{\rm B} = \frac{1}{d}\sum_{j=1}^d \PP[W_j \neq \wh W_j]$ and the convexity of divergence; \eqref{eq:WdXdBSC_2} uses the fact that $W_j$ is uniform on $\{-1,1\}$ and the data processing inequality for divergence; \eqref{eq:WdXdBSC_3} uses the fact that $W_j$'s are independent; \eqref{eq:WdXdBSC_4} follows from Corollary~\ref{co:W_disc}. Applying $h_2^{-1}$ to both sides, we get \eqref{eq:WdXdBSC_plb}. The lower bound \eqref{eq:WdXdBSC_blb} is a consequence of \eqref{eq:WdXdBSC_4}.
\end{IEEEproof}

\begin{figure}[h!]
\centering
  \includegraphics[scale = 0.9]{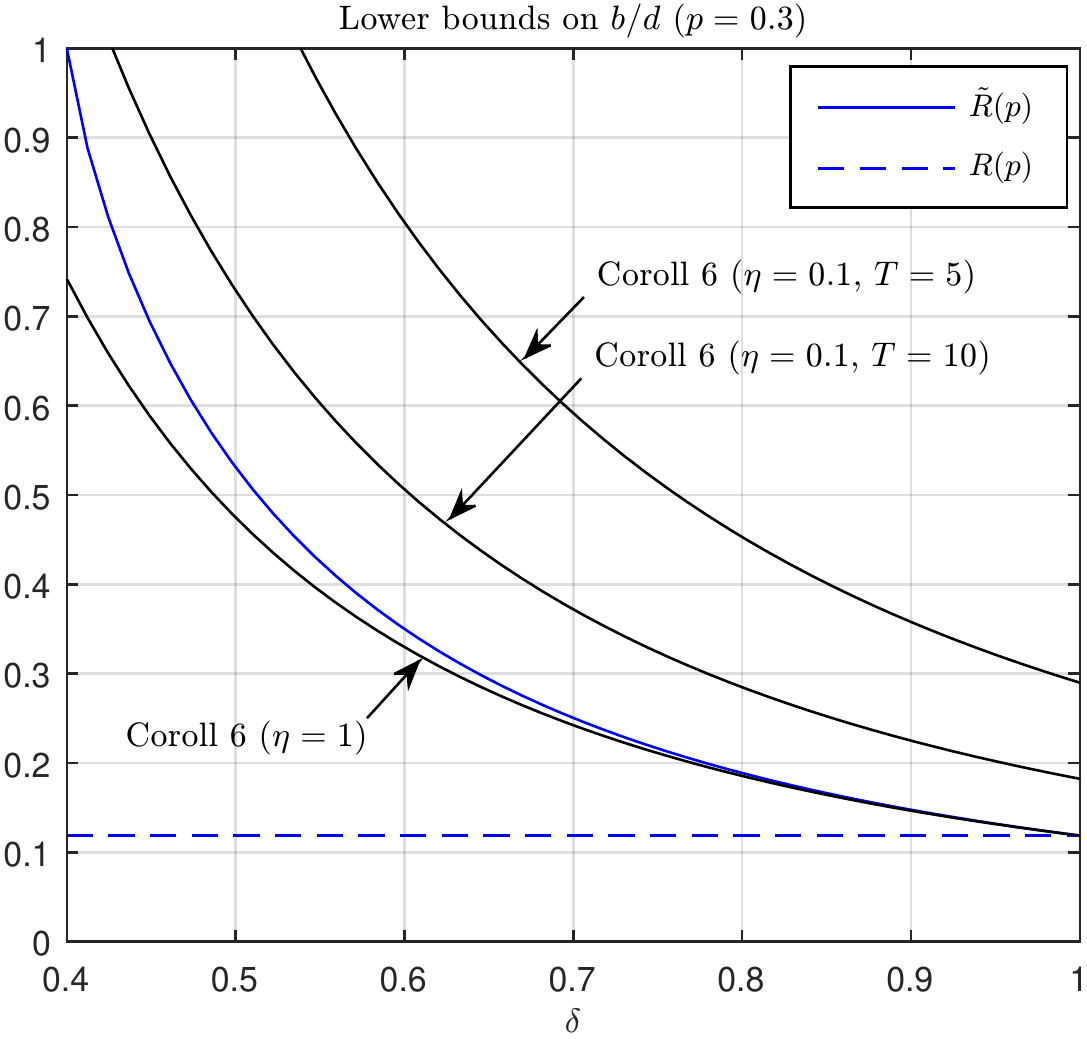}
  \caption{Comparison of lower bounds on $b/d$, where $p=0.3$ and $\eta = \eta(P_{V|U})$.}
  \label{fg:noisy_lossy}
\end{figure}

The asymptotic rate limit of noisy lossy coding of an i.i.d.\ ${\rm Bern}(\frac{1}{2})$ source observed through a ${\rm BSC}(\frac{1-\delta}{2})$ with distortion $p$ is given by
\begin{align}\label{eq:asymp_R_noisylossy}
\tilde R(p)= 1-h_2\left(\frac{2p+\delta-1}{2\delta}\right), \qquad 0\le \frac{1-\delta}{2}\le p \le \frac{1}{2} .
\end{align}
In Fig.~\ref{fg:noisy_lossy}, the lower bounds on the quantization rate given by \eqref{eq:WdXdBSC_blb} with different values of $\eta(P_{V|U})$ are compared with $\tilde R(p)$.
The lower bounds are also compared with the rate-distortion function of an i.i.d. ${\rm Bern}(\frac{1}{2})$ source, given by
\begin{align}
R(p)=1-h_2(p), \qquad 0\le p \le \frac{1}{2} .
\end{align}
We can see that with $\eta(P_{V|U}) = 1$, the lower bound well matches the asymptotically achievable rate given by \eqref{eq:asymp_R_noisylossy} for large $\delta$. With $\eta(P_{V|U}) < 1$, the elevated lower bounds capture the need to increase the quantization rate for sending the quantized message through another noisy channel.

\subsection{Estimating a continuous parameter}\label{sec:RB_cont}
\begin{example}\label{ex:W[01]_BSC}
Consider the problem of estimating the bias of a Bernoulli random variable through a BSC. In this case, $W$ is assumed to be uniformly distributed on $[0,1]$, $P_{X|W=w}$ is ${\rm Bern}(w)$, and $P_{V|U}$ is ${\rm BSC}(\eps)$. We are interested in lower-bounding the Bayes risk with respect to the absolute distortion $\ell(w,\wh{w}) = |w-\wh{w}|$.
\end{example}

Define $I^* = \sup_{\varphi_{\rm Q},\varphi_{\rm E}} I(W;V^T)$.
Replacing $I(W;X)$ with $I^*$ in \eqref{eq:RB_WU01_abs}, we obtain the following lower bound on the Bayes risk for this example as a consequence of Theorem~\ref{th:RB_lb_mi_diff}:
\begin{align}
R_{\rm B} \ge \frac{1}{2e} 2^{-I^*}  \label{eq:RB_cont_all_n} .
\end{align}
Now we only need to upper-bound $I^*$:

\begin{corollary}\label{co:I_cont}
In Example~\ref{ex:W[01]_BSC}, for any choice of $\varphi_{\rm Q}$ and $\varphi_{\rm E}$,
\begin{align*}
I(W;V^T) &\le \min\Big\{\Big(\frac{1}{2}\log n + \gamma_n\Big)\eta_T, \,\,
(1-2^{-n})b\eta_T, \,\,
(1-2^{-n})(1-h_2(\eps))T \Big\},
\end{align*}
where $\gamma_n$ is some sequence such that $\lim_{n\rightarrow\infty}\gamma_n = -0.6$, and $\eta_T = 1-(4\eps(1-\eps))^T$.
\end{corollary}
\begin{IEEEproof}
From \eqref{eq:I_WXn_ClaBar},
\begin{align}
I(W ; X^n) &= \frac{1}{2}\log\frac{n}{2\pi e} + h(W) + \frac{1}{2}\E\left[\log \frac{1}{W(1-W)}\right] + o(1) \\
&= \frac{1}{2}\log n -0.6 + o(1) \qquad \text{as $n \rightarrow \infty$}.
\end{align}
Moreover, from Lemma~\ref{lm:Dobru_coef},
\begin{align}\label{eq:Dbrsh_U_Bern}
\eta(P_{W|X^n}) \le \vartheta(P_{W|X^n}) = 1-2^{-n} ,
\end{align}
where the Dobrushin coefficient is evaluated in Appendix~\ref{appd:Dbrsh_U_Bern}. 
In addition, $\eta({\rm BSC}(\eps)) = (1-\eps)^2$.
With these facts, the result follows from Theorem~\ref{th:gen_single}.
\end{IEEEproof}
Now we apply the above results to two special cases.

\emph{Case 1}: $\eps=0$, $T\ge b$. 
In this case, the communication constraint only comes from the quantization of the samples, since the quantized message can be perfectly received by the estimator.
Setting $b=\tfrac{1}{2}\log n$, the lower bound in \eqref{eq:RB_cont_all_n} together with Corollary~\ref{co:I_cont} imply that
\begin{align}
R_{\rm B} &\ge \frac{1}{2e}2^{-(1-2^{-n})b} 
\ge \frac{1}{2e\sqrt{n}} .
\end{align}
To obtain an upper bound on $R_{\rm B}$, consider the scheme where the local processor computes the sample mean $\bar{X} = \frac{1}{n}\sum^n_{j=1}X_j$, which is uniformly distributed on $\{0,{1}/{n},\ldots,1\}$, and quantizes $\bar X$ into $\tilde X$ using a uniform $b$-bit quantization of $[0,1]$. The estimator sets $\wh W = \tilde X$. By the triangle inequality,
\begin{align}
\E|W-\wh W| &\le \E|W-\bar X| + \E|\bar X - \tilde X| 
\le \sqrt{\E[\Var(\bar X|W)]} + 2^{-b} \\
&= \frac{1}{\sqrt{6n}}+2^{-b}.
\end{align}
Thus for $b=\tfrac{1}{2}\log n$, 
\begin{align}
R_{\rm B} \le \frac{1.41}{\sqrt{n}} ,    
\end{align}
which differs from the lower bound only by a constant factor.

\smallskip

\noindent \textit{Case 2}: $\eps>0$, $b\ge\log(n+1)$. 
In this case, the communication constraint only comes from the noisy channel, since $\log(n+1)$ bits are enough to perfectly represent the sample mean $\bar X$, which is a sufficient statistic of $X^n$ for estimating $W$ and can take only $n+1$ values.
From \eqref{eq:RB_cont_all_n} and Corollary~\ref{co:I_cont}, 
\begin{align}\label{eq:BernBSC_RBlb}
R_{\rm B} &\ge \max\left\{\frac{1}{2e {n}^{\eta_T/2}}2^{-\gamma_n \eta_T}, \,\,
\frac{1}{2e} 2^{-(1-2^{-n})(1-h_2(\eps))T}\right\} .
\end{align}
To obtain an upper bound on $R_{\rm B}$, consider the scheme where the local processor first uses $\log(n+1)$ bits to represent the sample mean $\bar X$ as a message uniformly distributed on $\{0,1/n,\ldots,1\}$, then transmits the message over the channel using an optimal blocklength-$T$ code. The estimator decodes $\bar X$ as $\wh X$, and sets $\wh W = \wh X$. Then
\begin{align}
\E|W-\wh W| &\le \E|W - \bar X| + \E| \bar X - \wh X| \\ 
&\le \frac{1}{\sqrt{6n}} + \PP[\bar X \neq \wh X] \\ 
&\le \frac{1}{\sqrt{6n}} + 2^{-E_{\rm r}\left(\frac{1}{T}\log(n+1)\right)T} , \label{eq:BernBSC_RBub}
\end{align}
where $E_{\rm r}(\cdot)$ is the random coding error exponent of ${\rm BSC}(\eps)$ \cite[p.~146]{Gallager_ITbook}.
For $\frac{1}{T}\log(n+1) \le 1-h_2\big(\frac{\sqrt{\eps}}{\sqrt{\eps}+\sqrt{1-\eps}}\big)$,
\begin{align}
E_{\rm r}\big(\tfrac{1}{T}\log(n+1)\big) = 1 - \log(1+\sqrt{4\eps(1-\eps)}) - \tfrac{1}{T}\log(n+1) .
\end{align}
If the channel is used with feedback, then $E_{\rm r}(\cdot)$ in \eqref{eq:BernBSC_RBub} can be replaced by $E_{\rm f}(\cdot)$, the best attainable error exponent on BSC using block codes with feedback. In particular \cite[Problem 10.36]{CsiKor_book},
\begin{align}
\lim_{R\rightarrow 0}E_{\rm f}(R) = E_{\rm f}(0) = - \log\big(\eps^{1/3}(1-\eps)^{2/3} + \eps^{2/3}(1-\eps)^{1/3}\big) > E_{\rm r}(0) .
\end{align}
From the lower bound in \eqref{eq:BernBSC_RBlb} and the upper bound in \eqref{eq:BernBSC_RBub}, we know that the Bayes risk in this case decays polynomially in $n$ and exponentially in $T$.
Moreover,
\begin{align}
1\le \frac{1-h_2(\eps)}{- \log\big(\eps^{1/3}(1-\eps)^{2/3} + \eps^{2/3}(1-\eps)^{1/3}\big)} \le \frac{9}{8} \qquad \text{for $\eps\in\Big(\frac{2}{9},\frac{1}{2}\Big)$},
\end{align}
which implies that the error exponent with respect to $T$ in the lower bound can closely match that in the upper bound when transmission rate is low and $\eps$ is relatively large.

\section{Decentralized estimation: multiple processors}\label{sec:multi}
We now consider the problem setup with $m$ local processors. 
The $i$th processor, $i = 1,\ldots,m$, observes $n$ samples $X_{(i)}^n$ generated from a common random parameter $W$.
Given $W=w$, the joint distribution of the $m\times n$ array of samples is $P_{X_{(1)},\ldots,X_{(m)}|W=w}^{\otimes n}$.
In other words, the samples across different processors can be dependent conditional on $W$, but, at each processor $i$, the samples are i.i.d.\ draws from $P_{X_{(i)}|W=w}$.
As in the single-processor setup, the $i$th processor maps its samples to a $b$-bit message $Y_{(i)} = \varphi_{{\rm Q},i}(X_{(i)}^n)$, then maps the message to a blocklength-$T$ codeword $U_{(i)}^T = \varphi_{{\rm E},i}(Y_{(i)})$, and sends it to the estimator via $T$ uses of a discrete memoryless channel.
The estimator computes $\wh W = \psi(V^{m\times T})$ based on the received codewords $V^{m\times T} = (V_{(1)}^T,\ldots,V_{(m)}^T)$.
Here we assume that the channels between the processors and the estimator are independent and have the same probability transition law $P_{V|U}$.\footnote{The results can be straightforwardly generalized to the case where the parameters $n$, $b$, $T$, and the channels are different across the processors.}
The Bayes risk in this multi-processor setup is defined as 
\begin{align}
R_{\rm B} = \inf_{\varphi_{\rm Q}^m,\varphi_{\rm E}^m,\psi} \E\big[\ell(W,\psi(V^{m\times T}))\big] .
\end{align}

Compared with the single processor setup, the multi-processor setup gives rise to some new problems:
\begin{itemize}
\item
The sample sets observed by different processors can be either independent or dependent conditionally on $W$, depending on the joint distribution $P_{X_{(1)},\ldots,X_{(m)}|W=w}$.
In Section~\ref{sec:multi_ind_data}, we derive lower bounds for the case where $X_{(1)},\ldots,X_{(m)}$ are conditionally independent given $W$; in Section~\ref{sec:multi_dep_data}, we study the case where $X_{(1)},\ldots,X_{(m)}$ are dependent conditionally on $W$.
We will see that the Bayes risk can behave quite differently in these two cases.
\item
Suppose the $m\times n$ array of samples $(X_{(1)}^n,\ldots,X_{(m)}^n)$ can be observed by a single processor, which can map the samples to an $mb$-bit message and use the channel $mT$ times to send the message, and the estimation is based on the received codeword of blocklength $mT$.
How will the estimation performance degrade once these resources are distributed into $m$ processors in the multi-processor setup?
We examine this performance degradation through the Bayes risk lower bounds, for both cases where the sample sets are conditionally independent and dependent.
\item
When the channels are noiseless and feedback is available from the estimator to the local processors, each processor can observe the messages sent by the other processors. This allows for interactive protocols, as studied in \cite{Zhang_dist13,GMN14,BGMNW15}.
We will mainly focus on the case where the communication from local processors to the estimator is carried out without feedback, except for Section~\ref{sec:multi_inter}, where we consider the case where feedback is available and derive lower bounds on the Bayes risk for interactive protocols.
\end{itemize}

Before delving into various special cases, we give two general lower bounds for Bayes risk in the multi-processor setup, which are immediate consequences of Theorems~\ref{th:RB_lb_mi} and \ref{th:RB_lb_mi_diff} respectively:
\begin{theorem}\label{th:RBlb_multi_gen}
In the multi-processor setup, the Bayes risk satisfies
\begin{align}\label{eq:RBlb_dep_cond}
R_{\rm B} &\ge \inf_{\varphi_{\rm Q}^m, \varphi_{\rm E}^m} \sup_{\cS \subset [m],\,\rho>0} \rho \left(1-\frac{I(W; V^{m\times T}|X_{\cS}^n)+1}{\log({1}/{\E[\cL_{W}(X_{\cS}^n,\rho)]})}\right) ,
\end{align}
where $X_{\cS}^n = (X_{(i)}^n)_{i\in\cS}$.
When $W\in\R^d$ and $\ell(w,\wh w) = \|w-\wh w\|^r$ for any norm $\|\cdot\|$ in $\R^d$ and any $r\ge 1$, 
\begin{align}\label{eq:RB_lb_multi_mi_diff}
R_{\rm B}
&\ge \inf_{\varphi_{\rm Q}^m, \varphi_{\rm E}^m} \sup_{\cS \subset [m]} \frac{d}{re} \left(V_d \Gamma\Big(1+\frac{d}{r}\Big)\right)^{-{r}/{d}} 2^{-(I(W;V^{m\times T}|X_{\cS}^n) - h(W|X_{\cS}^n))r/d} .
\end{align}
\end{theorem}
\noindent The proof of Theorem~\ref{th:RBlb_multi_gen} is inspired by the proof of the Slepian-Wolf converse for distributed almost-lossless source coding using the cutset argument \cite[Chap. 15.4]{Cover_book}: choose the auxiliary random variable $U = X_{\cS}^n$ in Theorems~\ref{th:RB_lb_mi} and \ref{th:RB_lb_mi_diff}, then optimize over $\cS$. 

\subsection{Sample sets conditionally independent given $W$}\label{sec:multi_ind_data}
We first study the case where the sets of samples observed by the processors are conditionally independent given the parameter $W$. In this case, we can simply choose $\cS = \varnothing$ in Theorem~\ref{th:RBlb_multi_gen} to obtain lower bounds on the Bayes risk.
To that end, we need an upper bound on $I(W;V^{m\times T})$ which is independent of $\varphi_{\rm Q}^m$ and $\varphi_{\rm E}^m$:
\begin{theorem}\label{th:multi_iid_miub}
In the multi-processor setup, where the samples observed by the processors are conditionally i.i.d.\ given $W$, for any choice of $\varphi_{\rm Q}^m$ and $\varphi_{\rm E}^m$,
\begin{align}\label{eq:miub_gen_multi_iid}
I(W;V^{m\times T})& \le \min\Big\{ 
I(W;X^{m\times n})\eta_{mT}, \,\,
\eta(P_{X^n},P_{W|X^n}) mb \eta_T, \,\,
\eta(P_{X^n},P_{W|X^n}) mCT \Big\} ,
\end{align}
where $\eta_T = \eta(P_{V|U}^{\otimes T})$.
The first upper bound can be replaced by $mI(W;X^n)\eta_T$.
\end{theorem}
\begin{IEEEproof} 
Applying SDPI to the Markov chain $W - X^{m \times n} -  U^{m \times T} -  V^{m\times T}$, we get the first upper bound in \eqref{eq:miub_gen_multi_iid}.
Due to the independence assumption, the codewords $V^T_{(1)},\ldots,V^T_{(m)}$ received by the estimator are conditionally independent given $W$. This implies that (see, e.g., \cite[Lemma~4]{Duchi_dist})
\begin{align}
I(W;V^{m\times T})\le \sum_{i=1}^m I(W;V^T_{(i)}) .
\end{align}
Using Theorem~\ref{th:gen_single} to upper-bound each term, we obtain the second and the third upper bound in \eqref{eq:miub_gen_multi_iid}, as well as an alternative $mI(W;X^n)\eta_T$ to the first upper bound.
\end{IEEEproof}

To capture the penalty of decentralization, consider the situation where a total number of $N$ conditionally i.i.d.\ samples are allocated to a single processor, which maps them to a $B$-bit message and uses the channel $L$ times to send the message. In this situation, Theorem~\ref{th:gen_single} gives the upper bound
\begin{align}\label{eq:miub_multi_iid_1st}
I(W;V^{L})& \le \min\Big\{
I(W;X^{N})\eta_{L}, \,\, 
\eta(P_{X^{N}},P_{W|X^{N}}) B \eta_L, \,\, 
\eta(P_{X^{N}},P_{W|X^{N}}) CL \Big\} .
\end{align}
Once these resources are evenly distributed to $m$ processors, so that each processor observes $N/m$ samples, maps then to a $B/m$-bit message and uses the channel $L/m$ times to send the message, Theorem~\ref{th:multi_iid_miub} implies that
\begin{align}\label{eq:miub_multi_iid_dist}
I(W;V^{m\times \frac{L}{m}})& \le \min\Big\{
I(W;X^{N})\eta_{L}, \,\,
\eta(P_{X^{N/m}},P_{W|X^{N/m}}) B \eta_{L/m}, \,\, 
\eta(P_{X^{N/m}},P_{W|X^{N/m}}) CL \Big\} ,
\end{align}
where the first upper bound can be replaced by $mI(W;X^{N/m})\eta_{L/m}$.
Comparing \eqref{eq:miub_multi_iid_dist} with \eqref{eq:miub_multi_iid_1st}, we see that the differences are in the SDPI constants $\eta(P_{X^{N/m}},P_{W|X^{N/m}})$ and $\eta_{L/m}$.
Since $W - X^{n} - X^{k}$ form a Markov chain whenever $k\le n$, Lemma~\ref{lm:SDPI_ub_ss} implies that $\eta(P_{X^{N/m}},P_{W|X^{N/m}})$ is decreasing in $m$.
For example, when $W \sim N(0,\sigma_W^2)$ and $X_i = W+Z_i$ with $Z_i$  drawn i.i.d.\ from $N(0,\sigma^2)$ for $i=1,\ldots,n$, we have $\eta(P_{\bar X},P_{W|\bar X}) = {n \sigma_W^2}/({n\sigma_W^2 + \sigma^2})$ by \eqref{eq:SDPI_GaussXY}. Then by Lemma~\ref{lm:SDPI_ub_ss} 
\begin{align}
\eta(P_{X^{N/m}},P_{W|X^{N/m}}) &\le 
\frac{\sigma_W^2 N/m}{\sigma_W^2 N/m + \sigma^2} \\
&\approx \frac{N}{m}\frac{\sigma_W^2}{\sigma^2} \qquad\text{when $\frac{\sigma_W^2}{\sigma^2}$ is small} .
\end{align}
Moreover, from \eqref{eq:SDPI_TthPV|U} we know that $\eta_{L/m}$ is decreasing in $m$ as well, and
\begin{align}
\eta_{L/m} \approx \frac{L}{m} \eta({P_{V|U}}) \qquad \text{when $\eta(P_{V|U})$ is small.}
\end{align}
Thus, when the processors observe sample sets that are conditionally independent given the parameter, the penalty of decentralization can be captured by the reduced SDPI constants. The resulting upper bound on $I(W;V^{m\times \frac{L}{m}})$ decreases as the resources are distributed to more processors.

To illustrate the effectiveness of Theorem~\ref{th:multi_iid_miub}, we first show an example of mean estimation in the $d$-dimensional Gaussian location model with a Gaussian prior:
\begin{example}\label{ex:dGLM_GPrior_dec}
Consider the decentralized estimation of $W \sim N(0,\sigma^2_W {\bf I}_d)$ with $m$ processors, where the samples are i.i.d.\ draws from $N(w,\sigma^2 {\bf I}_d)$ given $W=w$.
The distortion function is $\ell(w,\wh w) = \|w-\wh w\|_2^2$.
Suppose there are $N$ samples in total, a budget of $B$ bits for quantization, and $L$ available uses of the channels. These resources are evenly distributed to the $m$ processors.
\end{example}
Combining \eqref{eq:miub_multi_iid_dist} from Theorem~\ref{th:multi_iid_miub} and \eqref{eq:RB_lb_multi_mi_diff} in Theorem~\ref{th:RBlb_multi_gen}, we get the following Bayes risk lower bound for Example~\ref{ex:dGLM_GPrior_dec}:
\begin{corollary}\label{co:dGLM_GPrior_dec}
In Example~\ref{ex:dGLM_GPrior_dec}, the Bayes risk satisfies
\begin{align}
R_{\rm B} \ge d \sigma_W^2 
\max\left\{
\left(1+\frac{N\sigma_W^2}{\sigma^2}\right)^{-\eta_L}\!\!\!, \,\,
\exp\left(- \frac{N\sigma_W^2 \ln 4 }{N\sigma_W^2 + m\sigma^2} \frac{(B\eta_{L/m} \wedge CL)}{d}\right)
\right\}
\end{align}
where $\eta_L = \eta(P_{V|U}^{\otimes L})$.
\end{corollary}
The first lower bound captures the increase of the Bayes risk due to the noisy communication channels, as compared to the Bayes risk $\frac{d \sigma_W^2 }{1+{N\sigma_W^2}/{\sigma^2}}$ of the centralized estimation.
From the second lower bound, we can see the order increase of the Bayes risk when the samples and the communication resources are distributed to more processors.
When the communication channels are noiseless, the lower bound in Corollary~\ref{co:dGLM_GPrior_dec} reduces to
\begin{align}
R_{\rm B} \ge
\max\left\{
\frac{d \sigma_W^2 }{1+N\sigma_W^2/\sigma^2}, \,\,
d \sigma_W^2  \exp\left(- \frac{N\sigma_W^2 \ln 4 }{N\sigma_W^2 + m\sigma^2} \frac{B}{d}\right) 
\right\}.
\end{align}
It shows that, with noiseless communication channels, in order to achieve the same performance as in the centralized scenario, the total number of bits allocated for quantization needs to be at least
\begin{align}\label{eq:Blb_dGLM_GW}
B \ge \left(1+\frac{m\sigma^2}{N\sigma_W^2}\right) \frac{d}{2} \log\left(1+\frac{N\sigma_W^2}{\sigma^2}\right) .
\end{align}
Note that it is necessary to have $N\ge m$, since each processor should observe at least one sample. Whether the lower bound in \eqref{eq:Blb_dGLM_GW} is a sufficient condition for achieving the Bayes rate of centralized estimation is an open problem.

As a second example, we use Theorem~\ref{th:multi_iid_miub} to derive lower bounds on the minimax risk for a nonparametric estimation problem studied in \cite{Zhang_dist13}. Here we assume that the communication channels are noisy:
\begin{example}\label{ex:minmax_bdd[pm1]^d}
Consider the decentralized estimation of the mean of an unknown distribution $P$ on $\sX = [-1,1]^d$, where each processor $i \in[m]$ only observes a single independent sample $X_{(i)}$ drawn from $P$. 
We use $\mathcal P$ to denote the family of probability distributions on $[-1,1]^d$, and define $\theta(P) = \E_P[X]$ for a distribution $P\in\mathcal P$. 
The minimax risk of this example is defined as
\begin{align}
R_{\rm M} = \inf_{\varphi_{\rm Q}^m,\varphi_{\rm E}^m,\psi} \, \sup_{P\in \mathcal P} \E_P \|\theta(P)- \psi(V^{m\times T})\|_2^2,
\end{align}
where $\psi$ is an estimator of $\theta \in [-1,1]^d$.
\end{example}
\begin{corollary}\label{co:RM_lb_dist}
In Example~\ref{ex:minmax_bdd[pm1]^d}, the minimax risk satisfies
\begin{align}
R_{\rm M} > \frac{d}{5}\min\Big\{1,\, \frac{d}{m \min\{d \eta_T, \, b\eta_T,\,CT\}}\Big\}  ,
\end{align}
where $\eta_T = \eta(P_{V|U}^{\otimes T})$.
\end{corollary}
\begin{IEEEproof}
At a high level, the proof strategy follows that in \cite{Zhang_dist13} by reducing the minimax estimation problem to the Bayes estimation problem in Example~\ref{ex:W{pm1}^dBSC} of Section~\ref{sec:W{pm1}^dBSC}.
However, here we use the result of Corollary~\ref{co:W_disc_RBrate} instead of the distance-based Fano's inequality used in \cite{Zhang_dist13} to obtain a tighter lower bound. The lower bound will also be able to capture the influence of noisy channels between the processors and the estimator. 

Let $W$, $\delta$, and $P_{X_j|W_j}$ be defined as in Example~\ref{ex:W{pm1}^dBSC}. Conditional on $W=w$, each processor observes an independent copy of $X$, whose coordinates are drawn according to $P_{X_j|W_j=w_j}$ for $i=1,\ldots,d$. Hence $P_{X|W=w}\in\mathcal P$ for all $w\in\{-1,1\}^d$. 
Let $\theta_w \deq \theta(P_{X|W=w}) = \delta w$, then 
\begin{align}
\|\theta_w-\theta_{w'}\|^2 = 4\delta^2\ell_{\rm H}(w,w') ,
\end{align}
where $\ell_{\rm H}$ denotes the Hamming distance. 
Define
\begin{align}
R_{\rm B} &= \inf_{\varphi^m_1,\varphi^m_2}\inf_{\psi}\E[\ell_{\rm H}(W,\wh W)],
\end{align}
where the second infimum is over all estimators of $W \sim {\rm Unif}(\{-1,+1\}^d)$.
Then, for $0\le \delta \le 1$,
\begin{align}\label{eq:RM_RB_dist}
R_{\rm M} &\ge 4\delta^2 R_{\rm B} .
\end{align}
From the proof of Corollary~\ref{co:W_disc_RBrate} and Theorem~\ref{th:multi_iid_miub}, we have 
\begin{align}
1 - h_2(R_{\rm B}/d) &\le  \frac{1}{d}I(W ; V^{m\times T}) 
\le \frac{\delta^2 m}{d}\min\Big\{d \eta_T, \, b\eta_T,\,CT\Big\}, 
\end{align}
where we have replaced the first upper bound in Theorem~\ref{th:multi_iid_miub} with $mI(W;X)\eta_T$, and used the fact that $1-h_2((1-\delta)/2) \le \delta^2$.
Thus,
\begin{align}
R_{\rm M} &\ge 4\delta^2 d \, h_2^{-1} \left(1 - \frac{\delta^2 m}{d}\min\{d \eta_T, \, b\eta_T,\,CT\} \right) .
\end{align}
With $\delta^2 = \min\{1,{d}/({2 m \min\{d \eta_T, \, b\eta_T,\,CT\}})\}$, the quantity in the parentheses is at least $1/2$, and since $h_2^{-1}(1/2) > 1/10$, we obtain the desired result.
\end{IEEEproof}
When the communication channels are noiseless, Corollary~\ref{co:RM_lb_dist} reduces to
\begin{align}
R_{\rm M} > \frac{d}{5}\min\Big\{1,\, \frac{d}{m (d \wedge b)}\Big\}  ,
\end{align}
which recovers the lower bound in \cite[Proposition~2]{Zhang_dist13} and improves the multiplicative constant. The lower bound can be achieved within a constant factor when $b = d$, using a method described in \cite{Zhang_dist13}.

As the last example of this section, we apply Theorem~\ref{th:multi_iid_miub} to the case where the parameter is a vector of length $n$, and each component of the sample set is generated according to the corresponding component of the parameter.
\begin{example}[CEO problem with noisy channels]\label{ex:noisyCEO}
Suppose the unknown parameter now is a random sequence $W^n$, consisting of $n$ i.i.d. draws from some prior distribution $P_W$ on $\R^d$.
$X_{(1)},\ldots,X_{(m)}$ are assumed to be independent, but not necessarily identically distributed, conditional on $W$.
Given $W^n = w^n$, the $i$th processor observes the sample set $X_{(i)}^n$, whose $j$th component is independently drawn from $P_{X_{(i)}|W=w_j}$, for $j=1,\ldots,n$. 
The $i$th processor then maps $X_{(i)}^n$ to a $b_i$-bit message and encodes it for transmission via $T$ uses of a noisy channel $P_{V|U}$.
The estimator computes $\wh W^n$ from the $m$ received codewords as an estimate of $W^n$.
The distortion is measured by $\frac{1}{n}\sum_{j=1}^n \|w_j-\wh w_j\|^r$ with some norm $\|\cdot\|$ on $\R^d$ and some $r\ge 1$.
\end{example}
When the channels between the processors and the estimator are noiseless, Example~\ref{ex:noisyCEO} coincides with the CEO problem \cite{Ber_CEO}.
Courtade \cite{Court_SDPI13} worked out a lower bound on the sum rate of the CEO problem using SDPI. The following result is an extension of the result in \cite{Court_SDPI13} to the case where the channels between the processors and the estimator are noisy:
\begin{corollary}\label{co:noisyCEO}
For the CEO problem with noisy channels in Example~\ref{ex:noisyCEO}, if 
$\frac{1}{n}\sum_{j=1}^n \E \|W_j-\wh W_j\|^r \le \alpha$,
then the quantization rates $b_i/n$, $i=1,\ldots,m$, need to satisfy
\begin{align}\label{eq:noisyCEO}
\sum_{i=1}^m \frac{b_i}{n} \eta(P_{X_{(i)}},P_{W|X_{(i)}}) \eta_T 
\ge  h(W) - \log \left(V_d \Big(\frac{\alpha re}{d}\Big)^{d/r} \Gamma\Big(1 + \frac{d}{r}\Big) \right) ,
\end{align}
where $\eta_T = \eta(P_{V|U}^{\otimes T})$.
\end{corollary}
\begin{IEEEproof}
Since $X_{(1)}^n,\ldots,X_{(m)}^n$ are conditionally independent given $W^n$, Theorem~\ref{th:multi_iid_miub} gives
\begin{align}
I(W^n;\wh W^n) &\le \sum_{i=1}^m {b_i} \eta(P_{X_{(i)}^n},P_{W^n|X_{(i)}^n}) \eta_T \\
&= \sum_{i=1}^m {b_i} \eta(P_{X_{(i)}},P_{W|X_{(i)}}) \eta_T ,
\end{align}
where the second step follows from the independence among $(W_j,X_{(i),j})$'s for each fixed $i=1,\ldots,m$, and the tensorization property of the SDPI constant (Lemma~\ref{lm:SDPI_tensor}).

Now define
$$R_W(\alpha) = \inf_{P_{\wh W|W}:\,\E\|W-\wh W\|^r\le \alpha} I(W;\wh W)$$
and
$$R_{W^n}(\alpha) = \inf_{P_{\wh W^n|W^n}:\frac{1}{n}\sum_{j=1}^n \E \|W_j-\wh W_j\|^r \le \alpha} I(W^n;\wh W^n)$$
be the rate-distortion functions of $W$ and $W^n$ respectively.
We have
\begin{align}
I(W^n;\wh W^n) &\ge R_{W^n}(\alpha) \label{eq:CEO_lb_Rn} \\
&= n R_{W}(\alpha) \label{eq:CEO_lb_R1} \\
&\ge n \left( h(W) - \log \left(V_d \Big(\frac{\alpha re}{d}\Big)^{d/r} \Gamma\Big(1 + \frac{d}{r}\Big) \right)\right) \label{eq:CEO_lb_diff}
\end{align}
where \eqref{eq:CEO_lb_Rn} is because of the assumption that $\frac{1}{n}\sum_{j=1}^n \E \|W_j - \wh W_j\|^r \le \alpha$; \eqref{eq:CEO_lb_R1} uses the additivity property of the rate-distortion function under additive distortions; and \eqref{eq:CEO_lb_diff} is a consequence of \eqref{eq:RD_lb_diff_vol}.
The proof of \eqref{eq:noisyCEO} is completed by combining the upper and lower bounds on $I(W^n;\wh W^n)$.
\end{IEEEproof}

\subsection{Dependent sample sets}\label{sec:multi_dep_data}
Now we consider the situation where the processors observe dependent sample sets conditional on the parameter.
To obtain tight Bayes risk lower bounds, we need to choose a suitable conditioning subset $\cS$ in Theorem~\ref{th:RBlb_multi_gen}.
Once $\cS$ is chosen, we need to evaluate or upper-bound the expected conditional small ball probability $\E[\cL_{W}(X_{\cS}^n,\rho)]$ or the conditional differential entropy $h(W|X_\cS^n)$.
We also need to upper-bound $I(W;V^{m\times T}|X_{\cS}^n)$ regardless of the choice of $\varphi_{\rm Q}^m$ and $\varphi_{\rm E}^m$ .
Here we give a general upper bound on $I(W;V^{m\times T}|X_{\cS}^n)$, which holds regardless of whether or not the sample sets are conditionally independent  given $W$:
\begin{theorem}\label{th:multi_miub_cond}
In the multi-processor setup, for any choice of $\varphi_{\rm Q}^m$ and $\varphi_{\rm E}^m$, and for any $\cS \subset [m]$,
\begin{align}\label{eq:multi_miub_cond}
I(W;V^{m\times T}|X_{\cS}^n)& \le \min\Big\{
I(W;X_{\cS^c}^{n}|X_{\cS}^n) \eta_{|\cS^c|T}, \,\,
\eta(\cS) |\cS^c|b \eta_{|\cS^c|T}, \,\,
\eta(\cS) |\cS^c|CT
\Big\} ,
\end{align}
where $\cS^c = [m] \setminus \cS$, $\eta_{|\cS^c|T}  = \eta\big(P_{V|U}^{\otimes |\cS^c|T}\big)$, and
\begin{align}
\eta(\cS) = \sup_{x_\cS^n}\eta\big(P_{X_{\cS^c}^n|X_\cS^n = x_\cS^n},P_{W|X_{\cS^c}^n, X_{\cS}^n = x_\cS^n}\big) .
\end{align}
In particular, when the channels are noiseless, we have
\begin{align}\label{eq:multi_miub_cond_noiseless}
I(W;V^{m\times T}|X_{\cS}^n)& \le \min\Big\{
I(W;X_{\cS^c}^{n}|X_{\cS}^n) , \,\,
\eta(\cS) |\cS^c|b 
\Big\} .
\end{align}
\end{theorem}
\begin{IEEEproof}
Appendix~\ref{appd:multi_miub_cond}. 
\end{IEEEproof}
Theorem~\ref{th:multi_miub_cond} can be used to capture the penalty of decentralization when the sample sets are conditionally dependent.
Consider the situation where all of the $m$ sample sets $X_{(1)}^n,\ldots,X_{(m)}^n$ are observed by a single processor, which maps them to an $mb$-bit message and uses the channel $mT$ times to send the message. In this situation, we have the upper bound
\begin{align}\label{eq:multi_on1st_miub_cond}
I(W;V^{m T}|X_{\cS}^n)& \le \min\Big\{
I(W;X_{\cS^c}^{n}|X_{\cS}^n) \eta_{mT}, \,
\eta(\cS) mb \eta_{mT}, \, 
\eta(\cS) mCT
\Big\} 
\end{align}
(see Appendix~\ref{appd:multi_miub_cond} for the proof). In particular, when the channels are noiseless, we have
\begin{align}\label{eq:multi_on1st_miub_cond_noiseless}
I(W;V^{m T}|X_{\cS}^n)& \le \min\Big\{
I(W;X_{\cS^c}^{n}|X_{\cS}^n) , \,
\eta(\cS) mb 
\Big\} .
\end{align}
Comparing \eqref{eq:multi_miub_cond} with \eqref{eq:multi_on1st_miub_cond}, we can see that, when the sample sets are dependent conditionally on $W$, the penalty of decentralization can still be captured by the reduced upper bound on $I(W;V^{m T}|X_{\cS}^n)$. 
In particular, when the channels are noiseless, for a fixed $\cS$, the second upper bound in \eqref{eq:multi_miub_cond_noiseless} is only a $\frac{m-|\cS|}{m}$ fraction of the second upper bound in \eqref{eq:multi_on1st_miub_cond_noiseless}.
However, this does not mean that choosing $\cS$ as large as possible leads to the tightest lower bound on the Bayes risk. The reason is that a larger $\cS$ generally corresponds to a larger $\E[\cL_{W}(X_{\cS}^n,\rho)]$ or a smaller $h(W|X_{\cS}^n)$, which may offset the decrease of the upper bound on $I(W;V^{m T}|X_{\cS}^n)$. The optimal $\cS$ to choose thus depends on the specific problem.

We study two examples to illustrate the effectiveness of combining the upper bound on $I(W;V^{m T}|X_{\cS}^n)$ in Theorem~\ref{th:multi_miub_cond} with the lower bounds in Theorem~\ref{th:RBlb_multi_gen}.
For simplicity, we focus on the case where the communication channels are noiseless.
\begin{example}\label{ex:mod2_Bern_2proc}
Consider a two-processor case, where $W \sim U[0,1]$ and $X_1,X_2 \in \{0,1\}$.
The conditional distribution $P_{X_{(1)},X_{(2)}|W=w}$ is specified as $P_{X_{(1)},X_{(2)}|W=w}(0,0) = P_{X_{(1)},X_{(2)}|W=w}(1,1) = ({1-w})/{2}$, and $P_{X_{(1)},X_{(2)}|W=w}(0,1) = P_{X_{(1)},X_{(2)}|W=w}(1,0) = {w}/{2}$. 
Note that $X_1$ and $X_2$ are marginally independent of $W$, but are jointly dependent on $W$.
In the decentralized estimation, processor $i$ observes $X_{(i)}^n$ and maps the samples to a $b$-bit message. The estimator computes $\wh W$ based on the noiselessly received messages.
The distortion function is $\ell(w,\wh w) = |w-\wh w|$.
\end{example}
\noindent For this example, we can choose $\cS = \{2\}$, then use \eqref{eq:RB_lb_multi_mi_diff} in Theorem~\ref{th:RBlb_multi_gen} and \eqref{eq:multi_miub_cond_noiseless} in Theorem~\ref{th:multi_miub_cond} to obtain the following lower bound on the Bayes risk:
\begin{corollary}
In Example~\ref{ex:mod2_Bern_2proc}, the Bayes risk satisfies
\begin{align}
R_{B} 
&\ge  \frac{1}{2e} 2^{-(1-2^{-n})b} \label{eq:RB_dist_dep_2p} .
\end{align}
\end{corollary}
\begin{IEEEproof}
Since $X_{(2)}^n$ is independent of $W$, $h(W|X_{(2)}^n) = h(W) = 0$.
Moreover, since $X_{(1)}^n$ and $X_{(2)}^n$ are independent, and $Z^n = X_{(1)}^n \oplus X_{(2)}^n$ is a sufficient statistic of $X_{(1)}^n$ and $X_{(2)}^n$ for $W$,
\begin{align}
\eta(P_{X_{(1)}^n|X_{(2)}^n = x_{(2)}^n} , P_{W|X_{(1)}^n,X_{(2)}^n = x_{(2)}^n}) = 
\eta(P_{Z^n}, P_{W|Z^n}) \qquad \text{for all $x_{(2)}^n$} ,
\end{align}
where $Z_i$'s are i.i.d. ${\rm Bern}(1/2)$ and $P_{Z_i|W=w} = {\rm Bern}(w)$.
As shown in Appendix~\ref{appd:Dbrsh_U_Bern}, $\vartheta(P_{W|Z^n}) = 1-2^{-n}$. Thus
\begin{align}
\sup_{x_{(2)}^n} \eta(P_{X_{(1)}^n|X_{(2)}^n = x_{(2)}^n} , P_{W|X_{(1)}^n,X_{(2)}^n = x_{(2)}^n}) \le 1-2^{-n} .
\end{align}
Combining \eqref{eq:RB_lb_multi_mi_diff} in Theorem~\ref{th:RBlb_multi_gen} and \eqref{eq:multi_miub_cond_noiseless} in Theorem~\ref{th:multi_miub_cond}, we get
\begin{align}
R_{B} 
&\ge  \frac{1}{2e} 2^{-I(W;Y_{(1)},Y_{(2)}|X_{(2)}^n) + h(W|X_{(2)}^n)} \\
&\ge  \frac{1}{2e} 2^{-(1-2^{-n})b}  ,
\end{align}
which proves the claim.
\end{IEEEproof}
In the extremal case when Processor~1 does not send anything to the estimator, no matter how many bits Processor~2 can send to the estimator, the Bayes risk is lower-bounded by 
\begin{align}
R_{B} &\ge \dfrac{1}{2e} ,
\end{align}
which follows from \eqref{eq:RB_dist_dep_2p} by setting $b=0$. This conforms to the fact that $X_{(2)}^n$ is independent of $W$.
It shows that the communication constraint can have much more severe effects on the estimation performance when the sample sets are dependent conditionally on the parameter, as compared to the case where the processors can observe samples that are conditionally i.i.d.\ given the parameter.

The lower bound in \eqref{eq:RB_dist_dep_2p} may not be tight in general.
Setting $b=\tfrac{1}{2}\log n$, \eqref{eq:RB_dist_dep_2p} implies that 
\begin{align}
R_{\rm B} &\ge \frac{1}{2e\sqrt{n}} .
\end{align}
This lower bound would be achievable up to a constant factor when Processor~1 could observe both $X_{(1)}^n$ and $X_{(2)}^n$, in which case the problem is reduced to Example~\ref{ex:W[01]_BSC} with noiseless channel.
But it is unlikely to be achievable when the sample sets are distributed to the two processors.
A recent paper of El Gamal and Lai \cite{ElGLai15} studies the problem of decentralized minimum-variance unbiased estimation of $W$ based on observations quantized at the rate of  $b/n$. It is shown that Slepian--Wolf rates are not necessary to achieve the centralized estimation performance, but in their protocol $b$ needs to be proportional to $n$.
The optimal rate region for this decentralized estimation problem is still unknown.

Now we examine the penalty of decentralization.
First consider the situation where a single processor can observe both $X_{(1)}^n$ and $X_{(2)}^n$ and map them to a $2b$-bit message. 
In this situation, \eqref{eq:RB_lb_multi_mi_diff} in Theorem~\ref{th:RBlb_multi_gen} together with \eqref{eq:multi_on1st_miub_cond_noiseless} lead to
\begin{align}
R_{B} &\ge \frac{1}{2e} 2^{-(1-2^{-n})2b}  .
\end{align}
Choosing $2b = \frac{1}{2}\log n$, we have
\begin{align}\label{eq:multi_on1st_RB_noiseless}
R_{B} &\ge \frac{1}{2e\sqrt{n}} .
\end{align}
For achievability, the processor can compute the sufficient statistic $Z^n = X_{(1)}^n \oplus X_{(2)}^n$, where $Z_i$'s are i.i.d. ${\rm Bern}(w)$ given $W=w$, and use $\frac{1}{2}\log n$ bits to uniformly quantize the sample mean of $Z^n$ over $[0,1]$.
Following the same analysis as in Case 1 of Example~\ref{ex:W[01]_BSC}, we obtain
\begin{align}
R_{B} &\le \frac{1.41}{\sqrt{n}} .
\end{align}
Thus the lower bound \eqref{eq:multi_on1st_RB_noiseless} is tight up to a constant factor in this situation.
Once the sample sets and the $2b = \frac{1}{2}\log n$ bits are distributed to the two processors, it follows from \eqref{eq:RB_dist_dep_2p} that
\begin{align}
R_{B} &\ge \frac{1}{2e n^{1/4}}   .
\end{align}
Compared with \eqref{eq:multi_on1st_RB_noiseless}, we can see the order increase of the lower bound.
Therefore, although the Bayes risk lower bound given by \eqref{eq:RB_dist_dep_2p} may be conservative, it can already reflect the penalty of distributing the sample sets and the communication resources to two processors.

Example~\ref{ex:mod2_Bern_2proc} can be extended to the $m$-processor case:
\begin{example}\label{ex:mod2_Bern_mproc}
Consider the following conditional distribution of a length-$m$ binary vector $(X_{(1)},\ldots,X_{(m)})$ given $W$:
\begin{align}
P_{X_{(1)},\ldots,X_{(m)}|W=w}(x_{(1)},\ldots,x_{(m)}) = 
\begin{cases}
({1-w}){2^{-(m-1)}} , & \text{if $x_{(1)}\oplus \ldots \oplus x_{(m)} = 0$} \\
{w}{2^{-(m-1)}} , & \text{if $x_{(1)} \oplus \ldots \oplus x_{(m)} = 1$} 
\end{cases} .
\end{align}
The vector $(X_{(1)},\ldots,X_{(m)})$ has the property that any  $m-1$ or fewer of its coordinates are independent of $W$, while the entire vector is dependent on $W$.
Moreover, $Z = X_{(1)} \oplus \ldots \oplus X_{(m)}$ is ${\rm Bern}(w)$ conditional on $W=w$, and $Z$ is a sufficient statistic of $(X_{(1)},\ldots,X_{(m)})$ for estimating $W$.
In decentralized estimation, the $i$th processor observes $X_{(i)}^n$, $i=1,\ldots,m$, and maps its samples to a $b$-bit message. The estimator computes $\wh W$ based on the noiselessly received messages.
The distortion function is $\ell(w,\wh w) = |w-\wh w|$.
\end{example}

With $\cS = \{2,\ldots,m\}$, following a similar analysis as in Example~\ref{ex:mod2_Bern_2proc}, we can show that
\begin{align}
h(W|X_{\cS}^n) = h(W) = 0 ,
\end{align}
and 
\begin{align}
\sup_{x_{\cS}^n}\eta(P_{X_{(1)}^n|X_{\cS}^n = x_{\cS}^n} , P_{W|X_{(1)}^n,X_{\cS}^n = x_{\cS}^n}) 
\le 1-2^{-n}  .
\end{align}
Thus combining \eqref{eq:RB_lb_multi_mi_diff} in Theorem~\ref{th:RBlb_multi_gen} with Theorem~\ref{th:multi_miub_cond}, we get a lower bound on the Bayes risk in Example~\ref{ex:mod2_Bern_mproc}:
\begin{align}
R_{B} 
&\ge  \dfrac{1}{2e} 2^{-(1-2^{-n})b}  . \label{eq:RB_dist_dep_mp}
\end{align}

Again, we can examine the penalty of decentralization.
In the situation where a single processor can observe $(X_{(1)}^n,\ldots,X_{(m)}^n)$ and map them to a $mb$-bit message, it follows from 
\eqref{eq:RBlb_dep_cond} in Theorem~\ref{th:RBlb_multi_gen} and \eqref{eq:multi_on1st_miub_cond_noiseless} that
\begin{align}
R_{B} &\ge \dfrac{1}{2e} 2^{-(1-2^{-n})mb}  .
\end{align}
Choosing $mb = \frac{1}{2}\log n$, we have
\begin{align}\label{eq:mp_on1st_RB_noiseless}
R_{B} &\ge \frac{1}{2e\sqrt{n}}  ,
\end{align}
which is tight up to a constant factor.
Once the sample sets and the $mb = \frac{1}{2}\log n$ bits are distributed to the $m$ processors, it follows from \eqref{eq:RB_dist_dep_mp} that
\begin{align}
R_{B} &\ge \frac{1}{2e n^{1/(2m)}}  .
\end{align}
Compared with \eqref{eq:mp_on1st_RB_noiseless}, we can see the order increase of the lower bound as $m$ increases, which reflects the penalty of distributing the sample sets and the communication resources to more processors.

\subsection{Interactive protocols}\label{sec:multi_inter}
When the communications channels are noiseless and feedback is available from the estimator to the processors, each processor can observe the messages sent by the other processors. This allows for the interactive protocols, as studied in \cite{Zhang_dist13,GMN14,BGMNW15}.
Here we consider a case where the processors take turns to send messages to the estimator, and each processor transmits only once. The message sent by a processor can depend on the previous messages sent by other processors, and is noiselessly received by the estimator.
This serial interactive setup has also been considered by Shamir \cite{Shamir_dist14}.
\begin{theorem}\label{th:multi_inter}
Consider the multi-processor setup, where the processors observe sample sets $X_{(1)}^n,\ldots,X_{(m)}^n$ that are conditionally i.i.d.\ given $W$, and where the message sent by the $i$th processor is given by
\begin{align}
Y_{(i)} = \varphi_{i}(X_{(i)}^n,Y^{i-1}) , \qquad i = 1,\ldots,m .
\end{align}
If the backward channel $P_{X|W}$ satisfies
\begin{align}\label{eq:mp_int_llr_cond}
\frac{{\rm d}P_{X|W=w}}{{\rm d}P_{X|W=w'}}(x) \ge {\alpha},  \qquad\text{for all $x \in \sX$ and $w,w' \in \sW$}
\end{align}
for some constant $\alpha \in (0,1]$, then, for any choice of $\varphi^m$ and $\psi$,
\begin{align}\label{eq:miub_multi_inter}
I(W;Y^{m})& \le \min\Big\{
I(W;X^{m\times n}) , \, (1-\alpha^n) mb \Big\} .
\end{align}
In particular, the above upper bound holds in the non-interactive case as well.
\end{theorem}
\begin{IEEEproof} 
Appendix~\ref{appd:multi_inter}.
\end{IEEEproof}

\begin{figure}[h!]
\centering
  \includegraphics[scale = 1.0]{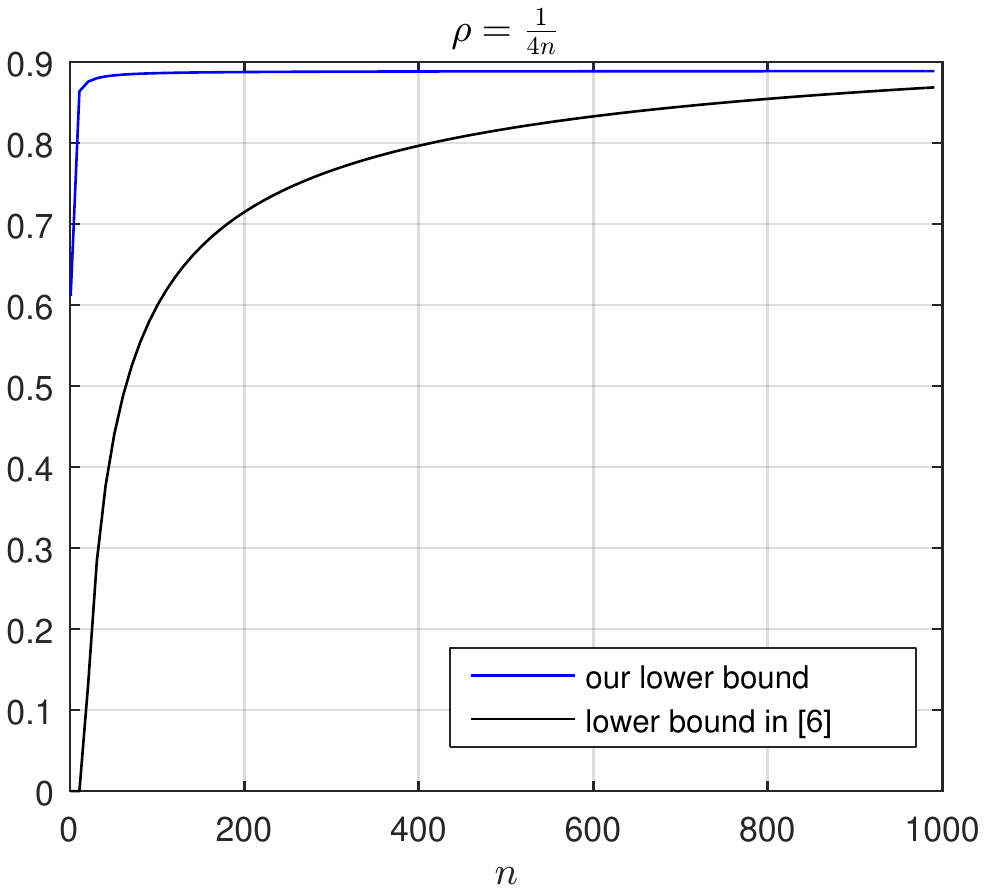}
  \caption{Comparison of minimax lower bounds given by Corollary~\ref{co:Rmlb_Shamir_HnS} and by \cite{Shamir_dist14}, where $m = 10$, $d=512$, $b=3d$, and $\rho = \frac{1}{4n}$.}
  \label{fg:Shamir_HnS_p14n}
\end{figure}

\begin{figure}[h!]
\centering
  \includegraphics[scale = 1.0]{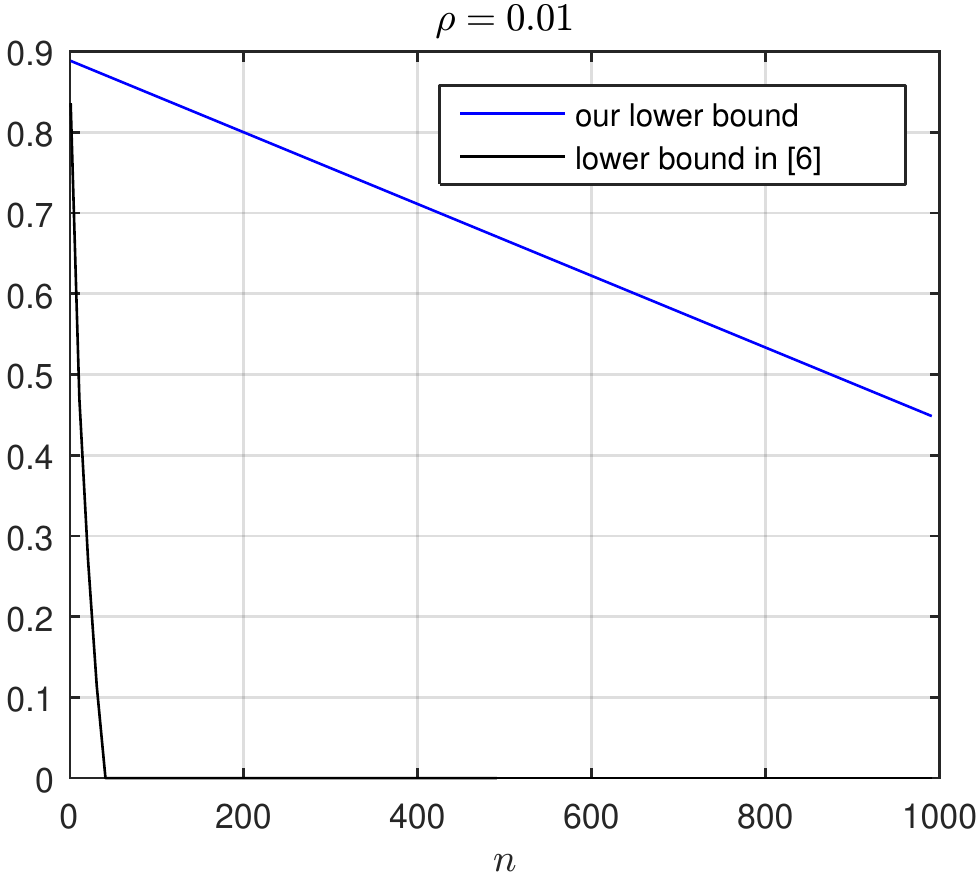}
  \caption{Comparison of minimax lower bounds given by Corollary~\ref{co:Rmlb_Shamir_HnS} and by \cite{Shamir_dist14}, where $m = 10$, $d=512$, $b=3d$, and $\rho = 0.01$ (the lower bound in \cite{Shamir_dist14} is set to $0$ when $n>1/4p$).}
  \label{fg:Shamir_HnS_p001}
\end{figure}

We can apply Theorem~\ref{th:multi_inter} to the ``hide-and-seek" problem formulated by Shamir \cite{Shamir_dist14} as a generic model for a number of decentralized estimation problems and online learning problems:
\begin{example}\label{ex:Shamir_HnS}
Consider a family of distributions $\mathcal P = \{P_w: w=1,\ldots,d\}$ on $\{0,1\}^d$.
Under $P_w$, the $w$th coordinate of the random vector $X \in \{0,1\}^d$ has bias $\frac{1}{2}+\rho$, while the other coordinates of $X$ are independently drawn from ${\rm Bern}(\frac{1}{2})$.
For $i=1,\ldots,m$, the $i$th processor observes $n$ samples $X_{(i)}^n$ drawn independently from $P_w$, and sends a $b$-bit message $Y_{(i)} = \varphi_{i}(X_{(i)}^n,Y^{i-1})$ to the estimator. The estimator computes $\wh W$ from the received messages $Y^m$.
The minimax risk of this example is defined as 
\begin{align}
R_{\rm M} = \inf_{\varphi^m,\psi} \, \max_{w\in [d]}\PP[\wh W \neq w] .
\end{align}
\end{example}
The minimax lower bound for this problem obtained in \cite{Shamir_dist14} is 
\begin{align}\label{eq:mxlb_shamir}
R_{\rm M} \ge 1 - \left(\frac{3}{d} + 5\sqrt{\min\left\{\frac{10\rho nmb}{d}, mn\rho^2\right\}}\right) \qquad\text{for $0\le \rho \le \frac{1}{4n}$} .
\end{align}
The question was left open whether this lower bound can be improved.
The following result gives an affirmative answer.
\begin{corollary}\label{co:Rmlb_Shamir_HnS}
In Example~\ref{ex:Shamir_HnS}, the minimax risk is lower bounded by
\begin{align}
R_{\rm M} \ge 1 - \frac{1}{\log d} 
\min\left\{
\left[1-\Big(\frac{1-2\rho}{1+2\rho}\Big)^n\right]mb + 1 ,\,
(4mn\rho^2 \wedge \log d) + 1
\right\} \qquad\text{for $0\le \rho \le \frac{1}{2}$} .
\end{align}
\end{corollary}
\begin{IEEEproof}
Let $W$ be uniformly distributed on $\{1,\ldots,d\}$. Then we can use the techniques developed so far to derive lower bounds on the average error probability $\PP[\wh W \neq W]$, which will provide lower bounds on the minimax risk. Using the fact that
\begin{align}
\frac{P_{X|W=w}(x)}{P_{X|W=w'}(x)} \ge \frac{\frac{1}{2}-\rho}{\frac{1}{2}+\rho} \qquad\text{for all $x \in \sX$ and $w,w' \in \sW$},
\end{align}
Theorem~\ref{th:multi_inter} gives
\begin{align}
I(W ; Y^m) &\le \left[1 - \Big(\frac{1-2\rho}{1+2\rho}\Big)^n\right] mb \qquad \text{for $0\le \rho \le \frac{1}{2}$} .
\end{align}
In addition, since the entries in $X^{m\times n}$ are i.i.d. conditional on $W=w$, defining $Q$ as the uniform distribution on $\{0,1\}^d$, we have
\begin{align}
I(W;X^{m\times n}) &\le mn D(P_{X|W} \| P_X | P_W) \\
&\le mn D(P_{X|W} \| Q | P_W) \label{eq:mi_gold_hidnsek}\\ 
&= mn\big(1-h_2(\tfrac{1}{2}+\rho)\big) \\ 
&\le 4mn\rho^2 
\end{align}
where \eqref{eq:mi_gold_hidnsek} follows from the identity $D(P_{X|W} \| P_{X} | P_W) = D(P_{X|W} \| Q | P_W) - D(P_X \| Q)$, and in the last step we have used the fact that $h_2(p) \ge 4p(1-p)$.
We also know that $I(W;X^{m\times n}) \le H(W) = \log d$.
Therefore, for $0\le \rho \le\frac{1}{2}$,
\begin{align}\label{eq:Shamir_HnS_miub}
I(W ; Y^m) &\le \min\left\{
\left[1-\Big(\frac{1-2\rho}{1+2\rho}\Big)^n\right]mb  ,\,
(4mn\rho^2 \wedge \log d) 
\right\} .
\end{align}
Moreover, the lower bound \eqref{eq:RBlb_dep_cond} in Theorem~\ref{th:RBlb_multi_gen} with the choice $\cS = \varnothing$ and the distortion function $\ell(w,\wh w) = \I\{\wh w \neq w\}$ becomes the usual Fano's inequality
\begin{align}
\PP[\wh W \neq W] \ge 1 - \frac{I(W;Y^m) + 1}{\log d} .
\end{align}
Plugging in the upper bound \eqref{eq:Shamir_HnS_miub}, we get the result.
\end{IEEEproof}
Now we compare the result of Corollary~\ref{co:Rmlb_Shamir_HnS} and the lower bound in \eqref{eq:mxlb_shamir}.
Note that the lower bound in \eqref{eq:mxlb_shamir} holds only for $0\le \rho\le \frac{1}{4n}$, whereas the lower bound given in Corollary~\ref{co:Rmlb_Shamir_HnS} holds for all $0\le\rho\le \frac{1}{2}$.
We compare them in two cases.
In the first case we set $\rho = \frac{1}{4n}$, and in the second case we set $\rho = 0.01$ for all $n$.
In both cases we set $m = 10$, $d = 512$, and $b = 3d$, as \cite{Shamir_dist14} considers the situation where $b = O(d)$.
With $n$ varying from $1$ to $1000$, we plot the lower bounds for the two cases in Fig.~\ref{fg:Shamir_HnS_p14n} and Fig.~\ref{fg:Shamir_HnS_p001} respectively.
We can see that the lower bound given by Corollary~\ref{co:Rmlb_Shamir_HnS} is tighter in the plotted range of $n$ in both cases.

\section{Conclusion}
We have proposed an information-theoretic framework for deriving general lower bounds on the Bayes risk in a systematic way, with  applications to  decentralized estimation.
The main contributions are summarized below.
\begin{itemize}
\item
Starting in the context of centralized estimation, we have derived lower bounds on the Bayes risk  in terms of mutual information (Theorem~\ref{th:RB_lb_mi}) and information density (Theorem~\ref{th:RB_sbp_NP}). 
Both lower bounds involve the small ball probability.
They are proved by lower-bounding the probability of excess distortion using properties of the Neyman-Pearson function, and then converting these bounds into lower bounds on the expected distortion using Markov's inequality.
The lower bounds in Theorem~\ref{th:RB_lb_mi} and Theorem~\ref{th:RB_sbp_NP} apply to general parameter spaces, prior distributions, sample generating models, and distortion functions.

\item
Theorem~\ref{th:RB_lb_mi_diff} gives a lower bound on the Bayes risk in terms of mutual information and differential entropy. The proof does not involve a detour to bounding the probability of excess distortion, and instead relies on the Shannon lower bound for the rate-distortion function, which directly relates the mutual information to the expected distortion.
Its unconditional version can yield tighter lower bounds than that of Theorem~\ref{th:RB_lb_mi}.
However, it only applies when the parameter space is $\R^d$ and the distortion is measured by some norm.

\item
All of our lower bounds on the Bayes risk  for centralized estimation involve an auxiliary conditioning random variable $U$.
A proper choice of $U$ can lead to tighter lower bounds than the ones without conditioning. Moreover, when applied to decentralized estimation, choosing $U$ as a subcollection of sample sets enables us to handle the case where the processors observe conditionally dependent sample sets (Theorem~\ref{th:RBlb_multi_gen}).

\item
In the context of decentralized estimation, the general results are refinements of the lower bounds on the Bayes risk based on mutual information (Theorem~\ref{th:RB_lb_mi} and Theorem~\ref{th:RB_lb_mi_diff}).
We have used strong data processing inequalities (SDPIs) as a unified method to quantify the contraction of mutual information caused by communication constraints.
The essence of this method is exhibited already in the upper bounds on the mutual information for the single-processor setup (Theorem~\ref{th:gen_single}).
For the multi-processor setup, we have discussed two cases depending on whether the sample sets are conditionally independent or not (Theorem~\ref{th:multi_iid_miub} and Theorem~\ref{th:multi_miub_cond}).
The resulting lower bounds on the Bayes risk (Theorem~\ref{th:RBlb_multi_gen}) provide us with a systematic way to quantify the penalty of decentralization.

\item
Finally, we have obtained upper bounds on the mutual information (Theorem~\ref{th:multi_inter}) for interactive communication protocols, where the processors take turns to send their messages, and each processor transmits only once.
Deriving general upper bounds on the mutual information using SDPIs for multi-round interactive protocols is an interesting direction for future research.

\end{itemize}

\section*{Acknowledgment} The authors would like to thank Yury~Polyanskiy and Yihong~Wu for helpful discussions and for making an early version of Ref.~\cite{PolWu_ES} available.
The authors also thank Thomas~Courtade for pointing out the connection to the noisy CEO problem.

\appendices

\renewcommand{\theequation}{\Alph{section}.\arabic{equation}}
\renewcommand{\thelemma}{\Alph{section}.\arabic{lemma}}
\setcounter{equation}{0}
\setcounter{lemma}{0}

\section{Proof of Lemma~\ref{lm:excess_lb_csbp} and Lemma~\ref{lm:excess_lb_iden}}\label{appd:RB_lb_NP}

The proof relies on the properties of the Neyman--Pearson function, which arises in the context of binary hypothesis testing, and is defined as follows: Given two probability measures $P$ and $Q$ on a common measruable space $\sZ$, for any $\alpha \in [0,1]$ let
\begin{align}\label{eq:def_NP}
\beta_\alpha(P,Q) = \inf_{f:\,\sZ\rightarrow [0,1]} \left\{ \int_\sZ f\, {\rm d} Q : \int_\sZ f\, {\rm d} P \ge \alpha \right\}.
\end{align}
We will need the following properties of $\beta_\alpha(P,Q)$:
\begin{itemize}
	\item Data processing inequality: For any Markov kernel $K$ from $\sZ$ into another measurable space $\sY$,
\begin{align}\label{eq:DP_NP}
	\beta_\alpha(PK, QK) \ge \beta_\alpha(P,Q),
\end{align}
where $PK$ and $QK$ are the images of $P$ and $Q$ under $K$ \cite{PolyanskiyVerdu10}.
\item Weak converse: For any $\alpha \in [0,1]$,
\begin{align}\label{eq:np_weak_conv}
d_2(\alpha \| \beta_\alpha) \le D(P \| Q) ,
\end{align}
where $d_2(p \| q) \deq p \log \frac{p}{q} + (1-p) \log \frac{1-p}{1-q}$ is the binary relative entropy \cite{PolWu_IT_lectures}.
\item Strong converse: For any $\alpha \in [0,1]$,
\begin{align}\label{eq:np_strong_conv}
\alpha - \gamma \beta_\alpha \le \left(1 - \gamma\inf_z \tfrac{{\rm d}Q}{{\rm d}P}(z)\right) P\left[\tfrac{{\rm d}P}{{\rm d}Q}(Z) \ge \gamma\right] \qquad\forall \gamma>0.
\end{align}
(see \cite[Lemma~35]{Pol_thesis}).
\end{itemize}
Now we proceed to the proof. Fixing an arbitrary $P_{U|W,X}$, define $\PP = P_{U,W,X}$ and $\QQ = P_U \otimes P_{W|U} \otimes P_{X|U}$. 
For any estimator $\psi:\sX\rightarrow\sW$ and any $\rho>0$, consider the function $f(w,x) = \I\{\ell(w,\wh w) < \rho\}$.
Then $\int f\, {\rm d} \PP = \PP[\ell(W,\wh W) < \rho]$ and $\int f\, {\rm d} \QQ = \QQ[\ell(W,\wh W) < \rho]$.
On the one hand,
\begin{align}
\QQ[\ell(W,\wh W) < \rho] &= \int_{\sU} \int_{\sW} \int_\sW \I\{\ell(w,\wh w) < \rho\} P_{W|U}({\rm d}w | u) P_{\wh W|U}({\rm d}\wh w | u) P_U({\rm d}u) \\
&= \int_{\sU} \int_{\sW} \PP[\ell(W,\wh w) < \rho | U = u] P_{\wh W|U}({\rm d}\wh w | u) P_U({\rm d}u) \\
&\le \int_{\sU} \sup_{\wh w\in\sW} \PP[\ell(W,\wh w) < \rho | U=u] P_U({\rm d}u) \\
&= \E[ \cL_{W|U}(U,\rho) ] . \label{eq:Qexcess_sbp}
\end{align}
On the other hand, by the definition of $\beta_\alpha$ and by the data processing inequality \eqref{eq:DP_NP},
\begin{align}
\QQ[\ell(W,\wh W) < \rho] 
&\ge \beta_{\PP[\ell(W,\wh W) < \rho]}\big(\PP_{W,\wh W}, \QQ_{W,\wh W}\big) \label{eq:whw_sbp_np} \\
&\ge \beta_{\PP[\ell(W,\wh W) < \rho]}(\PP, \QQ). \label{eq:sbp_np}
\end{align}
Combining \eqref{eq:Qexcess_sbp}, \eqref{eq:whw_sbp_np} and \eqref{eq:np_weak_conv}, and using the fact that $d_2(\alpha \| \beta) \ge \alpha\log\frac{1}{\beta} - h_2(\alpha)$, we obtain a lower bound on the excess distortion probability 
\begin{align}
\PP[\ell(W,\wh W) \ge \rho] \ge 1 - \frac{I(W;\wh W | U) + 1}{\log \big(1/\E[ \cL_{W|U}(U,\rho)] \big)} ,
\end{align}
which proves Lemma~\ref{lm:excess_lb_csbp}. 

Combining \eqref{eq:Qexcess_sbp}, \eqref{eq:sbp_np}, and \eqref{eq:np_strong_conv}, we obtain another lower bound on the excess distortion probability 
\begin{align}
\PP[\ell(W,\wh W) \ge \rho]  \ge  &\PP[i(W;X|U) < \log\gamma] - \gamma \E[{\mathcal L}_{W|U}(U,\rho)] + \nonumber \\
&\quad \gamma \inf_{u,w,x}\frac{{\rm d}P_{W|U=u}}{{\rm d}P_{W|U=u,X=x}}(w) \PP[i(W;X|U) \ge \log\gamma]  \qquad\forall \gamma>0 ,
\end{align}
which proves Lemma~\ref{lm:excess_lb_iden}.

\section{Proofs of Corollary~\ref{ex:GaussMean_GaussPrior} and Corollary~\ref{ex:WU[01]BernW}}\label{appd:GMean_GPrior_BernBias}
\subsection{Proof of Corollary~\ref{ex:GaussMean_GaussPrior}}
We prove this result using Theorem~\ref{th:RB_lb_mi}, by choosing $U$ as an conditionally independent copy of $X^n$ given $W$.
In Example~\ref{ex:GaussMean_GaussPrior}, we have the conditional pdf
\begin{align}
p_{W|X^n=x^n} = N\big(\E[W|X^n = x^n], \Var[W|X^n = x^n]\big) 
\end{align}
where
\begin{align}
\E[W|X^n = x^n] = \frac{\sigma_W^2}{\sigma_W^2 + {\sigma^2}/{n}} \bar{x}, \qquad 
\Var[W|X^n = x^n] = \frac{\sigma_W^2}{1+{n \sigma_W^2}/{\sigma^2}}  ,
\end{align}
and $\bar x = \frac{1}{n}\sum_{i=1}^n x_i$. Thus,
\begin{align}
\big\| p_{W|X^n=x^n} \big\|_{\infty} &= \sup_{w} |p_{W|X^n=x^n}(w)| = \sqrt{\frac{1}{2\pi} \left(\frac{1}{\sigma_W^2} + \frac{n}{\sigma^2}\right)} ,
\end{align}
and therefore
\begin{align}
\cL_{W|X^n}(x^n,\rho) &= \sup_{w \in \R} \PP[|W-w| < \rho | X^n = x^n] \\
&= \sup_{w \in \R} \int^{w+\rho}_{w-\rho} p_{W|X^n=x^n}(w') {\rm d} w' \\
&\le 2{\rho} \big\| p_{W|X^n=x^n} \big\|_{\infty}  \\
&= \rho\sqrt{\frac{2}{\pi} \left(\frac{1}{\sigma_W^2} + \frac{n}{\sigma^2}\right)}  .
\end{align}
In addition, 
\begin{align}
I(W ; X^n|X'^n) = I(W ; X^n, X'^n) - I(W ; X'^n) = \frac{1}{2}\log\frac{1 + 2n \sigma_W^2/\sigma^2}{1 + n\sigma_W^2/\sigma^2} .
\end{align}
From \eqref{eq:RB_sup_s}, 
\begin{align}
R_{\rm B} &\ge \sup_{0<s<1} \sqrt{\frac{\pi \sigma_W^2}{2(1+n\sigma_W^2/\sigma^2)}} s2^{-({I(W;X^n|X'^n) + 1})/{(1-s)}}  \\
&\ge  \frac{1 + \sigma^2/(n\sigma_W^2)}{8(2 + \sigma^2/(n\sigma_W^2))} \sqrt{\frac{\pi\sigma_W^2}{2(1+n\sigma_W^2/\sigma^2)}} \\
&\ge \frac{1}{16} \sqrt{\frac{\pi\sigma_W^2}{2(1+n\sigma_W^2/\sigma^2)}}  
\end{align}
where the second line follows by setting $s={1}/{2}$.

\subsection{Proof of Corollary~\ref{ex:WU[01]BernW}}
Again, we use Theorem~\ref{th:RB_lb_mi} by choosing $U$ as an conditionally independent copy of $X^n$ given $W$.
In Example~\ref{ex:WU[01]BernW}, we have the conditional pdf
\begin{align}
p_{W|X^n}(w|x^n) = (n+1) {n \choose k} (1-w)^{n-k} w^{k} \I\{0\le w \le 1\}
\end{align}
where $k = \sum_{i=1}^n x_{i}$. Since the maximum of the function $w \mapsto (1-w)^{n-k}w^k\I\{0 \le w \le 1\}$ is achieved at $w^* = k/n$, we have
\begin{align}
\big\| p_{W|X^n=x^n} \big\|_{\infty} = (n+1) {n \choose k} \Big(1-\frac{k}{n}\Big)^{n-k} \Big(\frac{k}{n}\Big)^{k} ,
\end{align}
and therefore
\begin{align}
\cL_{W|X^n}(x^n,\rho) 
&\le 2\rho \big\| p_{W|X^n=x^n} \big\|_{\infty}
= 2\rho (n+1) {n \choose k} \Big(1-\frac{k}{n}\Big)^{n-k} \Big(\frac{k}{n}\Big)^{k} .
\end{align}
Since the marginal distribution of $K=\sum_{i=1}^n X_i$ is uniform over $\{0,\ldots,n\}$, 
\begin{align}
\E[\cL_{W|X^n}(X^n,\rho)] &\le 2\rho \sum_{k=0}^n {n \choose k} \Big(1-\frac{k}{n}\Big)^{n-k} \Big(\frac{k}{n}\Big)^{k},
\end{align}
and, using Stirling's approximation \cite[p.~54]{Feller_vol1}, we have the estimate
\begin{align}
{n \choose k} \Big(1-\frac{k}{n}\Big)^{n-k} \Big(\frac{k}{n}\Big)^{k} \le \sqrt{\frac{n}{2\pi k(n-k)}} , \qquad k = 1,\ldots,n-1 .
\end{align}
With these upper bounds, we have
\begin{align}
\E[\cL_{W|X^n}(X^n,\rho)] &\le 2\rho \left(2 + \sum_{k=1}^{n-1}\sqrt{\frac{n}{2\pi k(n-k)}} \right) 
\le 2\rho \big(2 + \sqrt{{\pi n}/{2}}\big) .
\end{align}
In addition, from \eqref{eq:IWX|X'_asym},
\begin{align}
I(W ; X^n|X'^n) 
\rightarrow \frac{1}{2}  \qquad\text{as $n \rightarrow \infty$}.
\end{align}
Therefore, using Eq.~\eqref{eq:RB_sup_s},  we find
\begin{align}
R_{\rm B} &\ge \sup_{0<s<1} \frac{s}{2(2+\sqrt{\pi n/2})} 2^{-({I(W;X^n|X'^n) + 1})/{(1-s)}} \\
&\ge \frac{1}{4(2+\sqrt{\pi n/2})} 2^{-2 (I(W;X^n|X'^n) + 1)} \\
&\sim \frac{1}{16\sqrt{2\pi n}}  \qquad\text{as $n \rightarrow \infty$}  
\end{align}
where the second line follows by setting $s={1}/{2}$.

\section{Proof of Corollary~\ref{co:GaussMean_UPrior}}\label{appd:dGMean_UPrior}
We use the lower bound in \eqref{eq:RB_sbp_NP_uncond} to prove this result.
In Example~\ref{ex:GaussMean_UPrior}, the conditional pdf $p_{W|X^n=x^n}$ is a truncated Gaussian distribution
\begin{align}\label{eq:pW|X_uniball_Gauss}
p_{W|X^n}(w|x^n) &= \frac{\I\{\|w\|_2 \le a\}}{c_n(\bar{x})(2\pi\sigma^2/n)^{d/2}} \exp\left(-\frac{n}{2\sigma^2}\|\bar x - w\|_2^2\right) ,
\end{align}
where $\bar x = \frac{1}{n} \sum_{i=1}^n x_i \in \R^d$, and the normalizing factor is
\begin{align}
c_n(\bar{x}) &= \int_{\R^d} \frac{\I\{\|w\|_2 \le a\}}{(2\pi\sigma^2/n)^{d/2}} \exp\left(-\frac{n}{2\sigma^2}\|\bar x - w\|_2^2\right) {\rm d}w \\
&= \PP[\| \bar{X} + U_n  \|_2 \le a | \bar X = \bar x] 
\end{align}
with $U_n \sim N(0,\frac{\sigma^2}{n}{\mathbf I}_d)$ independent of $\bar{X}$.
We can show that\footnote{Given a sequence of real-valued random variables $\{Z_n\}$, we write $Z_n \xrightarrow{L^1} Z$, $Z_n \xrightarrow{P} Z$, and $Z_n \xrightarrow{d} Z$ to indicate the convergence in $L^1$, in probability, and in distribution, respectively.}
\begin{align}\label{eq:c_n_P}
c_n(\bar X) \xrightarrow{P} 1  \qquad \text{as $n\rightarrow \infty$}.
\end{align}
Indeed, since $\bar X \xrightarrow{P} W$ and $U_n \xrightarrow{d} 0$, we have $\bar X + U_n \xrightarrow{d} W$ \cite[Lemma~4.5 and Corollary 4.7]{Kallenberg}, hence
\begin{align}
\E[|c_n(\bar X) - 1|] &= 1 - \E[c_n(\bar X)] 
= \PP[\| \bar{X} + U_n  \|_2 > a] 
\rightarrow \PP[\|W\|_2>a] = 0  \qquad \text{as $n\rightarrow \infty$}
\end{align}
and thus $c_n(\bar{X}) \xrightarrow{L^1} 1$ as $n \to \infty$. Since $Z_n \xrightarrow{P} Z$ is equivalent to $\E[|Z_n-Z| \wedge 1] \to 0$ as $n \to \infty$, we arrive at \eqref{eq:c_n_P}. From \eqref{eq:pW|X_uniball_Gauss},
\begin{align}
\big\| p_{W|X^n=x^n} \big\|_{\infty} = 
\begin{cases}
\dfrac{1}{c_n(\bar{x})} \left(\dfrac{n}{2\pi\sigma^2}\right)^{d/2}, & \|\bar x\|_2 \le a \\
\dfrac{1}{c_n(\bar{x})} \left(\dfrac{n}{2\pi\sigma^2}\right)^{d/2} \exp\left(-\dfrac{n (\|\bar x\|_2-a)^2}{2\sigma^2}\right), & \|\bar{x}\|_2 > a 
\end{cases}.
\end{align}
Let $V_d$ denote the volume of the unit ball in $(\R^d, \| \cdot \|_2)$. Then, for all $x^n$ and $\|w\|_2\le a$,
\begin{align}
\frac{p_{W|X^n=x^n}(w)}{p_W(w)}
&\le V_d a^d \big\| p_{W|X^n=x^n} \big\|_{\infty} 
\le \frac{V_d a^d}{c_n(\bar{x})}\left(\frac{n}{2\pi\sigma^2}\right)^{d/2} .
\end{align}
Choosing $\gamma = (1 + \delta){V_d a^d}\left(\frac{n}{2\pi\sigma^2}\right)^{d/2} $ (for an arbitrary $\delta>0$) and $\rho = a(2\gamma)^{-1/d}$ in \eqref{eq:RB_sbp_NP_uncond}, we get
\begin{align}
R_{\rm B} &\ge  \rho \left(\PP\Big[i(W;X^n) < \log\gamma \Big] - \gamma {\mathcal L}_W(\rho)\right) \\
&\ge  \rho \left(\PP\left[\frac{V_d a^d}{c_n(\bar{X})}\left(\frac{n}{2\pi\sigma^2}\right)^{d/2} < \gamma \right] - \gamma \left(\frac{\rho}{a}\right)^d\right) \\
&\ge \Big(\frac{1}{2(1+\delta)}\Big)^{1/d} V_d^{-1/d} \sqrt{\frac{2\pi\sigma^2}{n}} \left(\PP\left[\frac{1}{c_n(\bar X)} < 1 + \delta \right] - \frac{1}{2}\right) \\
&\gtrsim \frac{1}{20} \sqrt{\frac{2\pi\sigma^2 d}{n}}  \qquad\text{as $n\rightarrow\infty$}
\end{align}
where the last step follows from the fact that $c_n(\bar{X}) \xrightarrow{P} 1$ (hence $1/c_n(\bar{X}) \xrightarrow{P} 1$), $(1/2)^{1/d}\ge 1/2$ for all $d\ge1$, $V_d^{1/d}\le {5}/{\sqrt{d}}$ for all $d\ge 1$, and the fact that $\delta>0$ is arbitrary.
We thus obtain a lower bound that is asymptotic in $n$ and non-asymptotic in $a$, $\sigma^2$, and $d$.

\section{Proof of (\ref{eq:Dbrsh_U_Bern})}\label{appd:Dbrsh_U_Bern}
We have $p_W(w) = 1$ for $w\in[0,1]$, and $P_{X^n|W}(x^n|w) = w^s(1-w)^{n-s}$, where $s$ is the Hamming weight (the number of $1$'s) of $x^n$. Thus,
\begin{align*}
P_{X^n}(x^n) = \int_0^1 w^s(1-w)^{n-s}{\rm d}w = \frac{1}{(n+1){n \choose s}}
\end{align*}
and
\begin{align*}
P_{W|X^n}(w|x^n) =  w^s(1-w)^{n-s}(n+1){n \choose s} .
\end{align*}
This gives
\begin{align*}
\|&P_{W|X^n=x^n} - P_{W|X^n=\tilde x^n}\|_{\rm TV} =  
\frac{n+1}{2}\int_0^1 \Big| w^s(1-w)^{n-s}{n \choose s} - w^{\tilde s} (1-w)^{n-\tilde s}{n \choose \tilde s} \Big| {\rm d}w ,
\end{align*}
which is maximized by choosing $x^n$ and $\tilde x^n$ such that $s = 0$ and $\tilde s = n$. Hence
\begin{align*}
\vartheta(P_{W|X^n}) &= \frac{n+1}{2}\int_0^1 \big| (1-w)^{n} - w^{n} \big| {\rm d}w  
= 1-2^{-n}.
\end{align*}

\section{Proofs of Theorem~\ref{th:multi_miub_cond} and Equation~\eqref{eq:multi_on1st_miub_cond}}\label{appd:multi_miub_cond}

\subsection{Proof of Theorem~\ref{th:multi_miub_cond}}
The first upper bound follows from
\begin{align}
I(W;V^{m\times T}|X_{\cS}^n) &= I(W;V_{\cS^c}^{T}|X_{\cS}^n) \label{eq:mi_WY1|X2_condMkv2} \\
&\le \eta(P_{V_{\cS^c}^{T}|U_{\cS^c}^{T}}) I(W;U_{\cS^c}^{T}|X_{\cS}^n) \label{eq:mi_WY1|X2_condMkv4} \\
&\le \eta_{|\cS^c|T} I(W ; Y_{\cS^c} | X_\cS^n) \label{eq:mi_WY1|X2} \\
&\le \eta_{|\cS^c|T} I(W ; X_{\cS^c}^n|X_{\cS}^n)
\end{align}
where 
\eqref{eq:mi_WY1|X2_condMkv2} follows from the Markov chain $W,V_{\cS^c}^{T} - X_{\cS}^n - V_{\cS}^{T}$,
and \eqref{eq:mi_WY1|X2_condMkv4} follows from the Markov chain $W,X_{\cS}^n - U_{\cS^c}^{T} - V_{\cS^c}^{T}$ and a conditional version of SDPI \cite[Lemma~1]{AXMR_dist_comp}.

Alternatively, we can upper-bound $I(W ; Y_{\cS^c} | X_\cS^n)$ in \eqref{eq:mi_WY1|X2} with the following chain of inequalities:
\begin{align}
I(W;V^{m\times T}|X_{\cS}^n)
&\le \eta_{|\cS^c|T} I(W ; Y_{\cS^c} | X_\cS^n)  \\
&= \eta_{|\cS^c|T} \int I(W ; Y_{\cS^c} | X_\cS^n = x_\cS^n) P_{X_\cS^n}({\rm d} x_\cS^n) \\
&\le \eta_{|\cS^c|T} \int I(X_{\cS^c}^n ; Y_{\cS^c} | X_\cS^n = x_\cS^n) \eta\big(P_{X_{\cS^c}^n|X_\cS^n = x_\cS^n},P_{W|X_{\cS^c}^n, X_{\cS}^n = x_\cS^n}\big) P_{X_\cS^n}({\rm d} x_\cS^n)  \label{eq:mi_WY1|X2_condMkv} \\
&\le \eta_{|\cS^c|T} \sup_{x_\cS^n} \eta\big(P_{X_{\cS^c}^n|X_\cS^n = x_\cS^n},P_{W|X_{\cS^c}^n, X_{\cS}^n = x_\cS^n}\big) |\cS^c| b , \label{eq:mi_WY1|X2_condMkv1}
\end{align}
where 
\eqref{eq:mi_WY1|X2_condMkv} is from the Markov chain $W - X_{\cS^c}^n - Y_{\cS^c}$ conditional on $X_\cS^n = x_\cS^n$ and the SDPI, and
\eqref{eq:mi_WY1|X2_condMkv1} is because $I(X_{\cS^c}^n ; Y_{\cS^c} | X_\cS^n) \le H(Y_{\cS^c}) \le |\cS^c|b$.

Lastly, from the Markov chain $W - X_{\cS^c}^n - V_{\cS^c}^T$ conditional on $X_\cS^n = x_\cS^n$ and the SDPI,
\begin{align}
I(W;V^{m\times T}|X_{\cS}^n) &= I(W;V_{\cS^c}^{T}|X_{\cS}^n) \\
&\le I(X_{\cS^c}^n;V_{\cS^c}^T | X_{\cS}^n) \sup_{x_\cS^n} \eta\big(P_{X_{\cS^c}^n|X_\cS^n = x_\cS^n},P_{W|X_{\cS^c}^n, X_{\cS}^n = x_\cS^n}\big) \\
&\le |\cS^c|CT \sup_{x_\cS^n} \eta\big(P_{X_{\cS^c}^n|X_\cS^n = x_\cS^n},P_{W|X_{\cS^c}^n, X_{\cS}^n = x_\cS^n}\big) ,
\end{align}
where the last step follows from $I(X_{\cS^c}^n;V_{\cS^c}^T | X_{\cS}^n) \le I(U_{\cS^c}^T;V_{\cS^c}^T | X_{\cS}^n) \le I(U_{\cS^c}^T;V_{\cS^c}^T)$, because of the Markov chain $X_{\cS}^n - U_{\cS^c}^T - V_{\cS^c}^T$.

\subsection{Proof of Equation~\eqref{eq:multi_on1st_miub_cond}}
The proof parallels that of Theorem~\ref{th:multi_miub_cond}.
For the first upper bound in \eqref{eq:multi_on1st_miub_cond},
\begin{align}
I(W;V^{mT}|X_{\cS}^n)
&\le \eta(P_{V^{mT}|U^{mT}}) I(W;U^{mT}|X_{\cS}^n) \label{eq:multi_on1st_miub_cond4} \\
&\le \eta_{mT} I(W ; Y | X_\cS^n) \label{eq:mi_on1st_WY1|X2} \\
&\le \eta_{mT} I(W ; X_{\cS^c}^n|X_{\cS}^n)  ,
\end{align}
where 
\eqref{eq:multi_on1st_miub_cond4} is from the Markov chain $W,X_{\cS}^n - U^{mT} - V^{mT}$.

Alternatively, we can upper-bound $I(W ; Y | X_\cS^n)$ in \eqref{eq:mi_on1st_WY1|X2} with the following chain of inequalities:
\begin{align}
I(W;V^{mT}|X_{\cS}^n)
&\le \eta_{mT} I(W ; Y | X_\cS^n) \\
&= \eta_{mT} \int I(W ; Y | X_\cS^n = x_\cS^n) P_{X_\cS^n}({\rm d} x_\cS^n) \\
&\le \eta_{mT} \int I(X_{\cS^c}^n ; Y | X_\cS^n = x_\cS^n) \eta\big(P_{X_{\cS^c}^n|X_\cS^n = x_\cS^n},P_{W|X_{\cS^c}^n, X_{\cS}^n = x_\cS^n}\big) P_{X_\cS^n}({\rm d} x_\cS^n)  \label{eq:multi_on1st_miub_cond5} \\
&\le \eta_{mT} \sup_{x_\cS^n} \eta\big(P_{X_{\cS^c}^n|X_\cS^n = x_\cS^n},P_{W|X_{\cS^c}^n, X_{\cS}^n = x_\cS^n}\big) m b , \label{eq:multi_on1st_miub_cond1}
\end{align}
where 
\eqref{eq:multi_on1st_miub_cond5} is from the Markov chain $W - X_{\cS^c}^n - Y$ conditional on $X_\cS^n = x_\cS^n$ and the SDPI;
\eqref{eq:multi_on1st_miub_cond1} is because $I(X_{\cS^c}^n ; Y | X_\cS^n) \le H(Y) \le mb$.

Lastly, from the Markov chain $W - X_{\cS^c}^n - V^{mT}$ conditional on $X_\cS^n = x_\cS^n$ and the SDPI,
\begin{align}
I(W;V^{m T}|X_{\cS}^n)
&\le I(X_{\cS^c}^n;V^{mT} | X_{\cS}^n) \sup_{x_\cS^n} \eta\big(P_{X_{\cS^c}^n|X_\cS^n = x_\cS^n},P_{W|X_{\cS^c}^n, X_{\cS}^n = x_\cS^n}\big) \\
&\le mCT \sup_{x_\cS^n} \eta\big(P_{X_{\cS^c}^n|X_\cS^n = x_\cS^n},P_{W|X_{\cS^c}^n, X_{\cS}^n = x_\cS^n}\big) ,
\end{align}
where the last step follows from $I(X_{\cS^c}^n;V^{mT} | X_{\cS}^n) \le I(U^{mT};V^{mT} | X_{\cS}^n) \le I(U^{mT};V^{mT})$, because of the Markov chain $X_{\cS}^n - U^{mT} - V^{mT}$.

\section{Proof of Theorem~\ref{th:multi_inter}}\label{appd:multi_inter}

The first upper bound in \eqref{eq:miub_multi_inter} follows from the Markov chain $W - X^{m\times n} - Y^m$.

\begin{figure}[h!]
\centering
  \includegraphics[scale = 1.1]{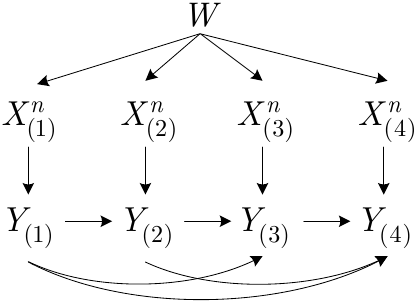}
  \caption{Bayesian network of $(W,X^{m\times n},Y^m)$ in the interactive case ($m=4$).}
  \label{fg:BN_WXY4}
\end{figure}

To prove the second upper bound in \eqref{eq:miub_multi_inter}, we use the chain rule to decompose $I(W;Y^m)$ as 
\begin{align}
I(W ; Y^m) &= \sum_{i=1}^m I(W : Y_{(i)}|Y^{i-1}) ,
\end{align}
and then apply SDPI to each term.
Since $Y_{(i)} = \varphi_{i}(X_{(i)}^n,Y^{i-1})$, we know that $W - X_{(i)}^n - Y_{(i)}$ form a Markov chain given $Y^{i-1} = y^{i-1}$.
Thus the SDPI gives
\begin{align}
I(W ; Y_{(i)} | Y^{i-1} = y^{i-1}) 
&\le \eta(P_{W | X_{(i)}^n, Y^{i-1} = y^{i-1}}) I(X_{(i)}^n ; Y_{(i)} | Y^{i-1} = y^{i-1}) .
\end{align}
Now the goal is to upper bound $\eta(P_{W | X_{(i)}^n, Y^{i-1} = y^{i-1}})$. 
We can view $P_{W | X_{(i)}^n, Y^{i-1} = y^{i-1}}$ as the backward channel and $P_{X_{(i)}^n | W, Y^{i-1} = y^{i-1}}$ as the forward channel.
Since we assume that each processor sends its message only once, $X_{(i)}^n$ and $Y^{i-1}$ are conditionally independent given $W$, which can be seen from the Bayesian network in Fig.~\ref{fg:BN_WXY4}.
Therefore,
\begin{align}
\frac{{\rm d}P_{X_{(i)}^n | W=w, Y^{i-1} = y^{i-1}}}{{\rm d}P_{X_{(i)}^n | W=w', Y^{i-1} = y^{i-1}}}(x_{(i)}^n) 
&= \frac{{\rm d}P_{X_{(i)}^n | W=w}}{{\rm d}P_{X_{(i)}^n | W=w'}}(x_{(i)}^n) \\
&\ge \alpha^n \quad\text{for all $x_{(i)}^n$, $w$, and $w'$} \label{eq:mp_int_llr_cond_n}
\end{align}
where \eqref{eq:mp_int_llr_cond_n} follows from the condition in \eqref{eq:mp_int_llr_cond} and the assumption that the samples in $X_{(i)}^n$ are conditionally i.i.d.\ given $W$.
Then by Lemma~\ref{lm:back_ch}, the SDPI constant of the backward channel satisfies
\begin{align}
\eta(P_{W | X_{(i)}^n, Y^{i-1} = y^{i-1}}) \le 1 - \alpha^n .
\end{align}
Since the above inequalities hold for any $y^{i-1}$, we have
\begin{align}
I(W ; Y_{(i)} | Y^{i-1}) &\le (1 - \alpha^n) I(X_{(i)}^n ; Y_{(i)} | Y^{i-1}) \\
&\le (1 - \alpha^n) I(X^{m\times n} ; Y_{(i)} | Y^{i-1}) .
\end{align}
It follows that
\begin{align}
I(W ; Y^m) &\le (1 - \alpha^n) I(X^{m\times n} ; Y^m) \\
&\le (1 - \alpha^n)mb .
\end{align}

\bibliography{distributed_estimation_journal_v5.bbl}

\end{document}